%% file: main.tex
\documentclass[amsmath,amssymb,aps,prl,twocolumn,superscriptaddress,nofootinbib]{revtex4-1}
\usepackage{graphicx,color,tikz,pgf,csquotes,hyperref}
\newcommand{\vtwo}[1]{#1} 

\newcommand{\cutoff}{L}

\newcommand{\tr}{\mathrm{tr}}
\def\d{\mathrm{d}}
\def\e{\textrm e}
\newcommand{\Z}{\mathbb Z}
\newcommand{\R}{\mathbb R}
\newcommand{\C}{\mathbb C}
\newcommand{\T}{\text{T}}

\newcommand{\SU}{\text{SU$(2)$}}
\newcommand{\SL}{\text{SL$(2,\C)$}}
\newcommand{\AL}{\text{A}^+_\cutoff}
\newcommand{\vol}[1]{V_{#1}}
\newcommand{\gf}{\Phi}
\newcommand{\cgf}{\Phi_0}
\newcommand{\gd}{{d_{\textsc{g}}}}
\newcommand{\ld}{{d_{\phi}}} 
\newcommand{\efd}{{d_{\textrm{eff}}}}
\newcommand{\crd}{{d_{\textrm{crit}}}}
\newcommand{\rk}{r}
\newcommand{\nv}{n}
\newcommand{\cs}{a}

\newcommand{\vg}{\boldsymbol{g}}
\newcommand{\rep}{j}
\newcommand{\vrep}{{\boldsymbol{\rep}}}
\newcommand{\ff}{\phi}       		
\newcommand{\vff}{\boldsymbol{\phi}}
\newcommand{\fm}{p}                 
\newcommand{\vfm}{\boldsymbol{p}}
\newcommand{\cas}{\textrm{Cas}} 

\newcommand{\nloph}{\hat{\mathcal{X}}}
\newcommand{\kin}{\mathcal{K}}

\renewcommand{\ell}{c}

\newcommand{\hyper}{\text{H}}

\newcommand{\Tr}{\mathrm{Tr}}

\begin{document}
\include{fdiagrams}

\title{Mean-field phase transitions in TGFT quantum gravity}

\author{Luca Marchetti}
\email{luca.marchetti@physik.lmu.de}
\affiliation{
Arnold Sommerfeld Center for Theoretical Physics,
Ludwig-Maximilians-Universit\"at M\"unchen,\\
Theresienstr. 37, 
80333 M\"unchen, Germany, EU
}
\author{Daniele Oriti}
\email{daniele.oriti@physik.lmu.de}
\affiliation{
Arnold Sommerfeld Center for Theoretical Physics,
Ludwig-Maximilians-Universit\"at M\"unchen,\\
Theresienstr. 37, 
80333 M\"unchen, Germany, EU
}
\affiliation{
Munich Center for Quantum Science and Technology (MCQST), Schellingstr. 4, 
80799 M\"unchen, Germany, EU
}
\author{Andreas G. A. Pithis}
\email{andreas.pithis@physik.lmu.de}
\affiliation{
Arnold Sommerfeld Center for Theoretical Physics,
Ludwig-Maximilians-Universit\"at M\"unchen,\\
Theresienstr. 37, 
80333 M\"unchen, Germany, EU
}
\affiliation{
Munich Center for Quantum Science and Technology (MCQST), Schellingstr. 4, 
80799 M\"unchen, Germany, EU
}
\author{Johannes Th\"urigen}
\email{johannes.thuerigen@uni-muenster.de}
\affiliation{
Mathematisches Institut der 
Westf\"alischen Wilhelms-Universit\"at, 
Einsteinstr. 62, 
48149 M\"unster, Germany, EU}
\affiliation{
Institut f\"ur Physik/Institut f\"ur Mathematik, 
Humboldt-Universit\"at zu Berlin, 
Unter den Linden 6, 
10099 Berlin, Germany, EU
}

\begin{abstract}
Controlling the continuum limit and extracting effective gravitational physics are shared challenges for quantum gravity approaches based on quantum discrete structures.
The description of quantum gravity in terms of tensorial group field theory (TGFT) has recently led to much progress in its application to phenomenology, in particular cosmology.
This application relies on the assumption of 
a phase transition to a nontrivial vacuum (condensate) state describable by mean-field theory, an assumption that is difficult to corroborate by a full renormalization group flow analysis due to the complexity of the relevant TGFT models.
Here we demonstrate that this assumption is justified due to the specific ingredients of realistic quantum geometric TGFT models: 
combinatorially non-local interactions, matter degrees of freedom and Lorentz group data together with the encoding of micro-causality.
This greatly strengthens the evidence for the existence of a meaningful continuum gravitational regime in group-field and spin-foam quantum gravity, the phenomenology of which is amenable to explicit computations in a mean-field approximation.

\end{abstract}


\maketitle


The main challenge for quantum gravity, in all its candidate formulations \cite{Blumenhagen:2013fgp,Percacci:2017fkn,Reuter:2019byg,Bonanno:2020bil,Ambjorn:2012jv,Hamber:2009zz,Surya:2019ndm,Rovelli:2004tv,Rovelli:2011eq,Perez:2012wv,Asante:2022dnj,Ashtekar:2004eh,Thiemann:2007pyv,Freidel:2005qe,Oriti:2011jm,Carrozza:2013oiy,Gurau:2016cjo,gurau2017random}, is to show the existence of a regime of the fundamental theory 
that reproduces gravitational physics together with new observable consequences.
This is highly non-trivial because we expect the new quantum gravity effects to become dominant at rather extreme energy and length scales. 
Still, early-universe cosmology, black holes and several astrophysical phenomena might provide many testing grounds for quantum gravity, also thanks to the recent improvement of observational techniques \cite{Addazi:2021xuf}. 
In quantum gravity formulations based on fundamentally discrete quantum structures, the challenge of extracting an effective continuum gravitational physics is intertwined with that of controlling the continuum limit of the fundamental quantum dynamics, i.e. its renormalization group flow and its continuum phase structure.

Black hole \cite{Oriti:2015rwa, Oriti:2018qty} and cosmological \cite{Gielen:2013cr,Pithis:2019tvp,Oriti:2021oux,Marchetti:2021gcv} physics 
have been studied in the framework of tensorial group field theory (TGFT) \cite{Freidel:2005qe,Oriti:2011jm,Carrozza:2013oiy}. 
This is a quantum field theory (QFT) generating spacetime geometries from discrete geometric building blocks given by combinatorially non-local interactions. 
This framework also provides a completion of the quantum dynamics encoded \vtwo{in spin foam models~\cite{Rovelli:2011eq,Perez:2012wv}, which is a covariant counterpart of canonical loop quantum gravity~\cite{Ashtekar:2004eh,Thiemann:2007pyv}
and a reformulation of specific \vtwo{simplicial} lattice gravity path integrals~\cite{Bonzom:2009hw,Baratin:2010wi,Baratin:2011tx,Baratin:2011hp,Finocchiaro:2018hks}.}

The underlying assumption in all these works is that there is a condensate phase, the dynamics of which can be captured by a mean-field approximation. However, this assumption is difficult to corroborate by a full renormalization group (RG) flow analysis due to the complexity of the relevant TGFT models. 
On the other hand, this 
complexity of the fundamental quantum dynamics is the result of 
incorporating the conditions required for a geometric interpretation at the discrete level, 
some seed of the causal structure expected to emerge in the continuum, 
and appropriate matter degrees of freedom. 

In this Letter we show that an appropriate mean-field description of a condensate phase 
exists in TGFT and that the quantum geometric and physical ingredients even improve the mean-field theory behaviour. 
To this aim, we apply a Landau-Ginzburg analysis, in particular the Ginzburg criterion, to models with 
(1) non-local interactions~\cite{Pithis:2018bw,Pithis:2019mlv}, 
(2) additional matter degrees of freedom \cite{Marchetti:2021xvf}, and
(3) the Lorentz group $\SL$,
together with the implementation of geometricity constraints \cite{Marchetti:2022igl}. 
\vtwo{For the first time we show how an effective dimension can be deduced from the Landau-Ginzburg setting;
as such notion relates to renormalization group results~\cite{Pithis:2020sxm,Pithis:2020kio} it allows us to argue for the validity of the result of two phases even beyond mean-field theory.}

Our results thus strongly support the existence of a meaningful continuum gravitational regime in TGFT quantum gravity as well as the closely related spin foam models and lattice quantum gravity.
\vtwo{This relies crucially on the completion of such discrete gravity models in terms of a sum over lattices generated by a field theory, thereby allowing to study their phenomenology by standard field-theory methods in a mean-field approximation.
In this sense}
the TGFT formulation provides new powerful tools for tackling the difficult problem of the continuum limit in lattice \vtwo{models of} quantum gravity and the quantum dynamics of spin network states. 

\section{Mean-field TGFT}

TGFT is a field theory that perturbatively generates a sum over lattices to which group-theoretic data are associated, encoding their discrete geometry \cite{Oriti:2011jm,Carrozza:2013oiy}.
The field $\gf:G^\rk\to\R$ excites $\rk$ elements $\vg=(g^1,...,g^\rk)$ in a Lie group $G$.
From the Feynman diagrams one obtains cell complexes due to combinatorially non-local interactions $\prod_i^n\gf(\vg_i)$,
\vtwo{i.e.~interactions in which the field arguments are convoluted} 
pairwise, one $g^a_i$ and one other $g^b_j$ respectively.
This is encoded by an $\rk$-valent interaction graph~$\gamma$ with $\nv_\gamma=n$ vertices and $\rk n/2$ edges $(i,a;j,b)$.
A generic action is thus 
\begin{equation}
S[\Phi] = \int_{G^\rk} \d\vg\, \Phi(\vg)\mathcal{K}\Phi(\vg) 
 +\sum_\gamma \frac{\lambda_\gamma}{\nv_\gamma} \Tr_\gamma\left[\Phi\right]
\end{equation}
where $\kin$ is a kinetic operator, 
$\sum_\gamma$ is over a given set of interaction graphs $\gamma$
and $\Tr_\gamma$ defines the resulting pairwise convolutions of each~$\gamma$ 
with kernels $\mathcal{V}$,
\begin{equation}\label{nonlocalinteraction}
\Tr_\gamma\left[\Phi\right]
= \int_{G^{\rk\cdot\nv_\gamma}} \prod_{i=1}^{\nv_\gamma} \d\vg_i \prod_{(i,a; j,b)} 
\mathcal{V}\left(g_i^a,g_j^b\right) 
\prod_{i=1}^{\nv_\gamma}\Phi(\vg_i) \, .
\end{equation}
If $a=b$ in each convolution, then the integers $a$ label the edges and one obtains an $\rk$-coloured graph (related to tensorial symmetry \cite{Bonzom:2012hw,gurau2017random}).
The resulting Feynman diagrams are then dual to $\rk$-dimensional simplicial pseudo-manifolds.

\vtwo{To associate discrete geometries with connection variables to the simplicial structures,} additional geo\-metricity constraints on the group elements called closure and simplicity constraints are necessary~\cite{Oriti:2011jm,Carrozza:2013oiy}.
With such constraints in place, the TGFT amplitude on a given simplicial lattice 
\vtwo{rewritten in terms of dual flux variables~\cite{Baratin:2010wi,Guedes:2013vi,Oriti:2018bwr}} takes the form of a lattice gravity path integral~\cite{Oriti:2011jm,Baratin:2011tx,Baratin:2011hp,Finocchiaro:2018hks}, 
\vtwo{or equivalently of a spin foam state sum~\cite{Perez:2012wv}.
There are different models distinguished by the specific implementation of such constraints, 
i.e.~a choice  of kinetic operator $\mathcal{K}$ and interaction convolutions~$\mathcal{V}$.
} 
TGFT thus differs significantly from local QFT in technical details and interpretation. 
Still, QFT methods can be adapted to this peculiar quantum gravity framework.

Mean-field theory provides an approximation of the full QFT partition function and thus an effective description of the phase structure~\cite{Kopietz:2010zz}.
One considers Gaussian fluctuations around a non-vanishing vacuum solution, e.g. for a constant mean-field~$\cgf$.
This is a \vtwo{
} solution to the TGFT equation of motion~\cite{Marchetti:2021xvf}
\begin{equation}\label{eq:eom}
\kin \gf + \sum_\gamma \frac{\lambda_\gamma}{\nv_\gamma} \sum_{v} \Tr_{\gamma\setminus v}(\gf) = 0 \, ,
\end{equation}
where the trace is over the graph $\gamma\setminus v$ obtained by deleting the vertex $v$ from $\gamma$
and $\sum_v$ runs over all its vertices $v$.

The validity of a mean-field description of phase transitions with diverging correlation length $\xi$, can be checked via the Ginzburg criterion. 
The ratio
\begin{equation}\label{eq:Q}
Q := \frac
{\int_{\Omega_\xi} \d\vg\, C(\vg) } 
{\int_{\Omega_\xi} \d\vg\, \cgf^2 }
\end{equation}
compares correlations $C(\vg)$ of Gaussian fluctuations with the vacuum $\cgf^2$ both averaged up to the correlation length scale~$\xi$, that is integrated over a suitable domain $\Omega_\xi$ \cite{Marchetti:2021xvf}.
If fluctuations remain small towards the phase transition, i.e.~$Q\ll1$ when $\xi\to\infty$, then mean-field theory is valid.
In local $d$-dimensional $\gf_d^4$ scalar field theory, 
$
Q \sim 
\vtwo{\mu^{-2}\xi^{-d}}
\sim\xi^{4 - d}
$
such that the mean-field description is valid only beyond the critical dimension $\crd=4$.
In the following, we explain how this is affected by the various features of TGFT models, progressively including all main ingredients for realistic quantum gravity models.

\section{1) Combinatorial non-locality}

The effect of combinatorially non-local interactions is most transparent for a simplified TGFT model. 
We choose 
$\kin=\sum_{\ell=1}^\rk -\Delta_\ell + \mu $ with $\Delta_\ell$ being the Laplacian on $G$
and Dirac delta convolutions $\mathcal{V} = \delta$.
Then the equation of motion \eqref{eq:eom} for the constant field $\cgf$ is
\begin{equation}\label{eqn:meanfield}
\bigg(\mu+\sum_\gamma \lambda_\gamma 
\vol{G}^{\rk\frac{\nv_\gamma-2}{2}} \cgf^{\nv_\gamma-2} \bigg)\cgf = 0 \, ,
\end{equation}
where $\vol G$ is the 
volume of $G$.
Such factors arise due to the combinatorial non-locality and need regularization if $G$ is not compact. 
For the example in this section we take $G=\R^\gd$ and regularize it to the $\gd$-torus $G_\cutoff=\T_\cutoff^\gd$ with radii $\cutoff/2\pi$ such that $\vol\cutoff \equiv\vol{G_\cutoff} =\cutoff^\gd$.

\begin{table}
    \centering
    \begin{tabular}{c|c|l}
        double trace & \cvft & $\nloph
         =4(2\prod\limits_{c=1}^4\delta_{\rep_c 0}+1)$
        \\ \hline
        $\nv=4$ melonic    & \cvf  & 
        $\nloph
        = 4(\prod\limits_{c}\delta_{\rep_c 0}+\prod\limits_{b\ne c}\delta_{\rep_b 0}+\delta_{\rep_c 0})$ 
        \\ \hline
        $\nv=4$ necklace & \cvfn & 
        $\nloph
        = 4(\prod\limits_{c}\delta_{\rep_c 0}+ \delta_{\rep_1 0}\delta_{\rep_2 0} +\delta_{\rep_3 0}\delta_{\rep_4 0})$
        \\ \hline
        simplicial
        & \cvfs &
        $\nloph
        = 5 \sum\limits_{i=0}^4 \prod\limits_{k\ne i} \delta_{\rep_{(ik)} 0}$ 
    \end{tabular} 
    \caption{Examples of non-local interaction graphs for $\rk=4$ and the resulting operator $\nloph_\gamma$ \cite{Marchetti:2021xvf}. 
    In addition to the usual legs (red) of interacting fields $\gf$, green half-edges represent the pairwise convolution of group arguments $g_i^a$ \cite{Thurigen:2021vy}.
    }
    \label{tab:vertexexamples}
\end{table}

For a single interaction $\gamma$, the vacuum equation \eqref{eqn:meanfield} reduces to
$(\vol\cutoff^{{\rk}/{2}}\cgf)^{\nv_\gamma-2} = -\frac{\mu}{\nv_\gamma\lambda_\gamma}$, which has a real solution for $\mu<0$.
The correlation function of Gaussian fluctuations around such $\cgf$ in momentum space is \cite{Marchetti:2021xvf}
\begin{equation}\label{eq:correlation}
\hat{C}(\vrep) 
= \frac{1 
}{\frac1{\vol\cutoff^2}\sum_c\cas_{\rep_c} + \mu - {\mu}\nloph_\gamma(\vrep)}
= \frac{1 
}{\frac1{\vol\cutoff^2}\sum_c\cas_{\rep_c} + b_\vrep} ,
\end{equation}
where $\rep_\ell$ labels representations of $G$, $\vrep\equiv (j_1,\dots,j_r)$ \vtwo{and $\cas_{\rep_c}$ denotes the Casimir of $\rep_c$}. 
On a \vtwo{compact or compactified} group these are countable, 
here $\rep_c\in\Z^\gd$.  
Moreover, $\nloph_\gamma(\vrep)$ is a sum over products of Kronecker deltas specific to
~$\gamma$ (see Table~\ref{tab:vertexexamples})
which gives rise to an effective mass $b_\vrep=\mu(1-\nloph_\gamma(\vrep))$.
Consequently, correlations expand in various multiplicities of zero modes,
\begin{equation}
C(\vg) = \frac{1}{\vol\cutoff^r} \sum_{s=0}^\rk \sum_{(c_1,..,c_s)} \sum_{\substack{\rep_{c_1},..,\rep_{c_s}=0\\ \rep_{c_{s+1}},..,\rep_{c_\rk}\neq 0 } } 
\!\!\!\tr_\vrep \bigg[ \hat{C}(\vrep)  \bigotimes_{\ell =1}^\rk D^{\rep_c}(g_c) \bigg],
\end{equation}
with representation matrices $D^{j_c}$, 
here 
$D^{j_c}(\e^{i\theta}) 
=\e^{i \theta\rep_c}$.
Since $D^0(g)=1$, each $s$-fold zero-mode $(c_1,...,c_s)$ contribution depends only on the other $\rk-s$ group variables
and the effective mass reduces to a number $b_\vrep = b_{c_1,..,c_s}$.

The correlation length $\xi$ can be obtained from the second moment or asymptotic behaviour of $C(\vg)$~\cite{Marchetti:2021xvf, Marchetti:2022igl}. 
It sets the characteristic scale beyond which correlations decay exponentially and diverges as $\xi^2 \sim -1/\mu$ at criticality, i.e.~when $\mu\to 0$.
For a large cut-off $\cutoff$, integrating $C(\vg)$ up to $\xi$ in each parameter $\theta_c^a$ yields \cite{Marchetti:2021xvf}
\begin{equation}
\int_{\Omega_\xi} \d\vg~C(\vg)
    = \sum_{s = s_0}^\rk \left(\frac{\xi}{\cutoff}\right)^{\gd s}
    \sum_{(c_1,...,c_s)}  \frac{1}{b_{c_1...c_s}}
\end{equation}
where $s_0$ is the minimal number of zero modes, that is the deltas in $\nloph$.
Interaction graphs $\gamma$ have multiple edges in general; 
it is the maximal multiplicity $s_\text{max}$ occurring in~$\gamma$ which determines $s_0$; in the model here $s_0=\rk-s_\text{max}$.

Taking the ratio \eqref{eq:Q} with integrated $\cgf^2$ gives 
\begin{align} \label{eq:Qpolynomial}
Q_\cutoff &=\left(\frac{
\lambda_\gamma}{-\mu}\right)^{\frac{2}{\nv_\gamma-2}} \sum_{s=s_{0}}^\rk \frac{f^\gamma_s}{-\mu} \left(\frac{\xi}{\cutoff}\right)^{-\gd (\rk-s)}
\end{align}
where we abbreviate the coefficients in the polynomial
$f^\gamma_s:=-\mu\sum_{(c_1,.., c_s)} b_{c_1,..,c_s}^{-1}$.
Removing the cut-off $\cutoff$, only the $s_0$-fold zero modes of the interaction $\gamma$ survive, yielding large-$\xi$ asymptotics (using $-\mu\sim \xi^{-2}$) 
\begin{align}\label{eq:Qnoncompact}
Q 
&\underset{\xi\to\infty}{\sim} 
\bar{\lambda}_{\gamma}^{\frac{2}{\nv_\gamma-2}}  f^\gamma_{s_0} \xi^{\frac{2\nv_\gamma}{\nv_\gamma-2}-\gd(\rk-s_0)}
\end{align}
with $\bar{\lambda}_\gamma = {\cutoff^{\gd(\rk-s_0)(\nv_\gamma-2)/2}}\lambda_\gamma$.
This is the result $Q\sim \vtwo{\mu^{-\frac{\crd}{2}}\xi^{-d}} \sim \xi^{\crd-d}$ of local QFT with 
$\crd = \frac{2\nv_\gamma}{\nv_\gamma-2}$.
The effect of the non-local interaction given by the graph $\gamma$ with $s_0$ minimal zero modes is a reduction of the configuration space dimension $\gd\rk$ to an effective dimension
\begin{equation}\label{eq:effdimensionflat}
d = \efd := \gd(\rk-s_0)\, .
\end{equation}

A typical model of 4d quantum gravity has $\rk=4$, group dimension 
at least $\gd=3$ (e.g. $G=\SU$ models), and quartic, quintic, or higher-order interactions.
Though the configuration space is then at least 12-dimensional and $\crd=4$ or $10/3$, respectively, the effective dimension $\efd$ might be smaller, e.g. for~$\rk-s_0=1$ in the case of $\gamma$ with no multiple edges like the simplicial interaction (see Table~\ref{tab:vertexexamples}).
Adding gauge invariance can shift $s_0 \to s_0+1$ and thus reduce $\efd$ even further \cite{Marchetti:2021xvf}. 
On the other hand, if there are edges of high enough multi\-plicities, e.g. melonic interactions with $s_0=1$,
$\efd$ is larger than $\crd$ such that $Q\ll 1$ and mean-field theory is a valid description of phase transitions.
Thus, combinatorial non-locality of TGFT affects the detailed mean-field behaviour but does not spoil the very applicability of mean-field theory.

A special case includes models where $G$ is compact and no thermodynamic limit is applicable. 
Then, there is a fixed finite volume $L^\gd$ (no $\cutoff\to \infty$ limit)
and it is the $\rk$-fold zero mode which dominates
in the IR, i.e., at small momentum scale $k\sim1/\xi$ \cite{Pithis:2020kio}.
In the current language, this is the $s=\rk$ mode,
and thus $\efd=0$.
In other words, also in TGFT one has ~\cite{Carrozza:2016tih, Pithis:2020sxm} the standard result that a QFT on a compact domain is effectively zero dimensional in the IR and does not allow for phase transitions \cite{Benedetti1403}. 

\section{2) Matter degrees of freedom}

Adding matter degrees of freedom in a TGFT model increases the effective dimension \cite{Marchetti:2021xvf}. 
We extend the group field to $\gf(\vff,\vg)$ with $\ld$ local degrees of freedom $\vff = (\ff_1,...,\ff_{\ld})\in\R^\ld$ with point-like interactions $\int\d\vff\,\Tr_\gamma\left[\Phi\right](\vff) 
$ while the convolutions of $\vg_i$ remain as in Eq.~\eqref{nonlocalinteraction}.
This generates free scalar fields minimally coupled to the discrete geometry.
Such fields can be used as reference frames which allows to retrieve the dynamics of quantum geometry in relational terms~\cite{Oriti:2016qtz,Li:2017uao,Gielen:2018fqv,Marchetti:2020umh}.
Furthermore, this is 
a necessary ingredient of realistic models of quantum gravity coupled to elementary particle fields.

In the Landau-Ginzburg analysis such local degrees of freedom simply add to the dimension.
Gaussian correlations \eqref{eq:correlation} extend to
\begin{equation}
\hat{C}(\vfm,\vrep) 
= \frac{1 
}{\alpha(\vrep) \sum_{a=1}^\ld \fm_a^2 + \frac1{\vol\cutoff^2}\sum_c\cas_{\rep_c=1}^\gd + b_\vrep}
\end{equation}
where the momenta $\fm_a$ of $\ff_a$ can couple to group representations $\vrep$ via a function $\alpha(\vrep)$.
However, only modes with $\fm_a\approx 0$ are relevant upon integration over $\ff_a$, so 
the Ginzburg~$Q$, Eq~\eqref{eq:Qnoncompact}, receives no contribution from $\alpha(\vrep)$ and  simply gets an additional factor $\xi^{-\ld}$ \cite{Marchetti:2021xvf}.
Accordingly, the effective dimension is
\begin{equation}
\efd = \ld + \gd(\rk-s_0) .
\end{equation}
Therefore, even if $\gd(\rk-s_0)\le\crd$, the Ginzburg criterion is still satisfied when there are $\ld>\crd-\gd(\rk-s_0)$ matter fields.
Even in the special case of compact $G$ without the thermodynamic large-volume limit, 
mean-field phase transitions are possible if $\ld>\crd$.
In this way, adding such matter degrees of freedom not only makes these models more realistic from a physical point of view, but also improves their mean-field behaviour.

\section{3) Hyperbolic geometry}
Realistic models of quantum gravity involve a Lie group with hyperbolic geometry related to the Lorentz group.
The latter is a crucial ingredient to properly implement micro-causality.

One such model is the Lorentzian Barrett-Crane TGFT model~\cite{Barrett:1999qw,Perez:2000ec,Perez:2000ep,Jercher:2021bie,Jercher:2022mky} which generates triangulations formed by space-like tetrahedra. 
This is a TGFT on the non-compact curved group $G=\SL$ subject to closure and simplicity constraints. 
Due to the restriction to space-like tetrahedra the group domain is effectively $\rk=4$ copies of the $3$-hyperboloid $\mathrm{H}^3=\SL/\SU$. 
Owing to the interplay with combinatorial non-locality, infinite volume factors do not cancel and need to be regularized (see e.g.\ \eqref{eqn:meanfield}). 
We implement this in the Cartan decomposition $\SL \cong \SU \times \text{A}^+ \times \SU$ via a cut-off $\cutoff$ on the diagonal Cartan subgroup
\begin{equation}
\text{A}^+= \{ \e^{\frac{\sigma_3}{2}\frac{\eta}{\cs}} \vert \eta\in \R_{+}\}
\to 
\AL := \{ \e^{\frac{\sigma_3}{2}\frac{\eta}{\cs}} \vert 0\le\eta<\cutoff\} \, .
\end{equation}
Here $\cs$ is the curvature scale, i.e.~the skirt radius of the group's hyperbolic part $\hyper^3$.
Using the Haar measure on $\SU$ and $\sinh^2(\eta/\cs) \d\eta/\cs$ for $\text{A}^+$, one finds the regularized volume of $\SL$ to be 
\begin{equation}\label{eq:volume}
\vol{\cutoff} 
= \frac{1}{4}\left(\sinh\left(\frac{2\cutoff}{\cs}\right) - \frac{2\cutoff}{\cs} \right)
\underset{\cutoff\to\infty}{\sim} \frac{1}{8}\e^{\frac{2\cutoff}{a}}\,.
\end{equation} 

The relative magnitude of Gaussian fluctuations around the mean-field $\cgf$ at a large cut-off $\cutoff$ is \cite{Marchetti:2022igl}
\begin{equation}\label{eq:Qhyperbolicallmodes}
Q_\cutoff = \frac{
\sum_{s=s_0}^\rk \frac{f^\gamma_s }{-\mu}  \vol\xi^s \, \vol\cutoff^{1-s} 
}{
\left(\frac{-\mu}{\lambda_\gamma}\right)^{\frac{2}{\nv_\gamma-2}} \vol\xi^\rk \, \vol\cutoff^{1-\rk-2\frac{\nv_\gamma-1}{\nv_\gamma-2}}
}\, .
\end{equation}
Removing the regularization, the minimal number $s_0$ of zero modes dominates such that 
\begin{align}\label{eq:Qhyperbolic}
    Q_\cutoff  &\underset{\cutoff\to\infty}\sim
    \bar{\lambda}_\gamma^{\frac{2}{\nv_\gamma-2}} 
    (-\mu)^{-\frac{\nv_\gamma}{\nv_\gamma-2}} 
    f^\gamma_{s_0} \vol\xi^{-(\rk-s_0)} 
\end{align}
with rescaling
$
\bar{\lambda}_\gamma =    
\vol\cutoff^{\frac{\nv_\gamma-2}{2}\left(\rk-s_0\right)+\nv_\gamma-1}
\lambda_\gamma
$.
Due to the hyperbolic geometry of the configuration space, the correlation length scales 
$\xi \underset{\mu\to 0}{\sim} \frac{1}{\cs \mu}$ \cite{Marchetti:2022igl}.
Thus, 
\begin{equation}\label{eq:Qhyperbolicscaling}
Q \underset{\xi\to\infty}\sim
\bar{\lambda}_\gamma^{\frac{2}{\nv_\gamma-2}} f^\gamma_{s_0}
(\cs\xi)^\frac{\nv_\gamma}{\nv_\gamma-2} 
\e^{-2(\rk-s_0)\frac{\xi}{\cs}}.  
\end{equation}
This is 
an exponential suppression such that the Ginzburg criterion $Q 
\ll 1$ is satisfied regardless of the order $\nv_\gamma$ and zero modes $s_0$ of the interaction. 

\begin{figure}
\centering
\includegraphics[width=.4\textwidth]{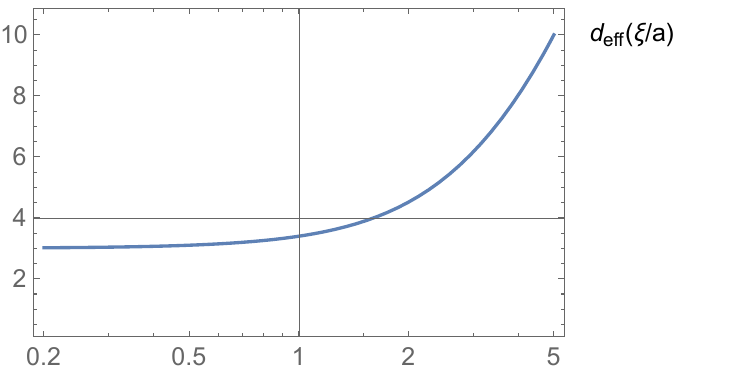}
\caption{Effective dimension \eqref{eq:effdimensionhyperbolic} for $s_0=\rk-1$ flowing from~$3$ at small $\xi$ to infinity.}
\label{fig:effdimension}
\end{figure}

A complementary way to understand this is to use the notion of a scale-dependent effective dimension $\efd$ \cite{Pithis:2020kio}.
One can still write Eq.~\eqref{eq:Qhyperbolic} as a power function but with scale-dependent inverse power
\begin{equation}\label{eq:effdimension}
\efd(\xi) := -\frac{ \partial \log F(\xi)}{\partial \log\xi}
\text{ , here }
F(\xi) = f^\gamma_{s_0} \vol{\xi}^{-(\rk-s_0)} \, .
\end{equation}
Then, the general form $Q \propto \mu^{-\crd/2}\xi^{-\efd(\xi)}$ is 
valid on all scales $\xi$, with the 
resulting effective dimension 
\begin{equation}\label{eq:effdimensionhyperbolic}
\efd(\xi) = (\rk-s_0)\frac{\cosh(\frac{2\xi}{\cs})-1}{\frac{\cs}{2\xi} \sinh(\frac{2\xi}{\cs})-1}
\end{equation}
flowing from $\efd=3(\rk-s_0)$ in the UV (small $\xi$)%
\footnote{Small $\xi$ is equivalent to large curvature $\cs$ here.
Thus, this limit also shows consistency with the flat case in Eq.~\eqref{eq:effdimensionflat} with $\gd=3$ \cite{Marchetti:2022igl}
because only the hyperbolic part $\mathrm{H}^3$ contributes here.
}
to $\efd\to\infty$ in the IR, see Figure~\ref{fig:effdimension}.
Thus, mean-field theory can describe the phase transition because the theory's effective dimension blows up towards the IR and thereby supersedes any possible critical dimension. 

This result for the Barrett-Crane TGFT model suggests that mean-field theory is valid for any TGFT on hyperbolic spaces 
irrespective of the dimension $\gd\cdot\rk$  of the field domain or of combinatorics with $\nv_\gamma$ and $s_0$.
Including $\ld$~matter fields adds again a power $\xi^{-\ld}$ \cite{Marchetti:2022igl} but the exponential suppression still dominates.
A hyperbolic sector also arises for models with compact $G$ but Lorentzian integrals implicit in the kernel $\kin$~\cite{Finocchiaro2012}. Similarly, one expects this exponential decay to remain also for TGFT models 
that also contain light- and time-like tetrahedra like in the complete Lorentzian Barrett-Crane model \cite{Jercher:2022mky}.
Thus, it seems to be a generic feature of TGFT quantum gravity that phase transitions towards a \vtwo{non-perturbative} vacuum state exist which can be self-consistently described using mean-field theory. 
Since such a state is typically highly populated by TGFT quanta, this indeed makes a compelling case for 
an interesting continuum geometric approximation. 

\vtwo{A reasonable viewpoint on this general behavior is that of universality of the continuum limit as naturally expected from coarse-graining, implemented here via mean-field techniques. This resonates with recent results on effective spin foam models which suggest that models differing in the precise implementation of the above-mentioned geometricity constraints could lie in the same universality class from the perspective of continuum gravitational physics~\cite{Asante:2020qpa,Asante:2020iwm,Dittrich:2021kzs,Asante:2021zzh,Dittrich:2022yoo}.}

\section{Conclusions}

The key result of this work is that it is possible to understand some part of the phase structure of TGFT with straightforward QFT methods. 
The full theory space is very involved and largely out of reach for explicit control due to the intricate interplay between 
the tensorial nature of the field, 
the Lie group domain and its geometry, 
combinatorially non-local interactions, as well as  
geometricity constraints.
Still, considering Gaussian fluctuations around a constant background vacuum field, we find that these features of TGFT can be controlled and even work in favour of a mean-field description:
(1) The general $\xi^{\crd-d}$ asymptotic scaling towards the IR remains even with non-locality, only with a modified effective dimension $d=\efd$.
(2) Coupling local degrees of freedom on~$\R^\ld$ adds to the dimension $\efd \to \efd+\ld$.
Most importantly, (3) the holonomies of the Lorentz group produce an exponential suppression of the fluctuation size 
such that we can always find a transition towards a phase 
which is self-consistently described in terms of mean-field theory regardless of the critical dimension. 

Crucially, this provides \vtwo{evidence for the existence of a condensate phase in quantum geometric TGFTs, the mean-field hydrodynamics of which can be mapped to an effective continuum cosmological dynamics~\cite{Oriti:2016qtz,Jercher:2021bie}. In this way, our work gives} evidence for the existence of a sensible continuum limit
in TGFT quantum gravity as well as in the closely related lattice quantum gravity and spin foam models where renormalization and the continuum limit are the main outstanding challenges~\vtwo{\cite{Bahr:2016hwc,Asante:2020qpa,Asante:2020iwm,Steinhaus:2020lgb,Dittrich:2021kzs,Asante:2021zzh,Dittrich:2022yoo,Asante:2022dnj}. Importantly, this is directly rooted} in the Lorentz group and thus the causal structure it enforces, which is reminiscent of results of the causal dynamical triangulations approach \cite{Ambjorn:2004qm,Ambjorn:2012jv,Jordan:2013awa,Ambjorn:2020rcn,Ambjorn:2022naa}. 
\vtwo{Overall, our results relate to and impact on a wider} set of quantum gravity approaches based on discrete structures comprising lattice quantum gravity and loop quantum gravity.
 
The exponential-suppression effect allows for an educated guess of the phase structure even beyond the Gaussian approximation.
Functional renormalization group calculations of local QFT on $\mathrm{H}^3$ in the local-potential approximation have shown that the flowing effective potential freezes around the curvature scale $\cs$~\cite{Benedetti1403} such that the entire phase diagram is already described by the two phases of the mean-field regime. 
Using the floating-point method one can identify the Wilson-Fisher fixed point but it is pushed to infinity when $(k a)^{-1}\sim\xi/a \to \infty$, where $k$ denotes the RG scale~\cite{Benedetti1403}.
This agrees with our result that
~$\efd$ flows to infinity in this regime such that the non-trivial phase structure at finite dimension vanishes.
\vtwo{Widening the scope from the mean-field perspective applied here towards that of the functional renormalization of TGFT}, we can thus also expect%
\footnote{
Higher orders in a derivative expansion such as the anomalous dimension may uncover further structure as in Ref.~\cite{Pithis:2020kio}.
But all fixed points found occur only up to some dimension.
Thus, we expect them to vanish when $\efd$ blows up in the IR. 
}
that hyperbolic geometry yields a universal phase structure already captured by the two phases around the Gaussian fixed point, 
\vtwo{though numerical details might still depend on the specific model chosen.}

By and large, our results greatly strengthen the evidence for the existence of a continuum regime in TGFT quantum gravity, the phenomenology of which is tractable by explicit computations in a mean-field approximation. 
This is all the more important given how difficult it is, due to their analytic and combinatorial complexity, to establish the same fact via a complete RG analysis of the relevant quantum geometric TGFT (and spin foam) models.

\

\begin{acknowledgments}
The work of L.~Marchetti was funded by Fondazione Angelo Della Riccia. 
D.~Oriti and A.~Pithis acknowledge funding from DFG research grants OR432/3-1 and OR432/4-1.
A. Pithis is grateful for the generous financial support from the MCQST via the seed funding Aost 862983-4.
J.~Th\"{u}rigen's research was funded by DFG grant number 418838388 and Germany's Excellence Strategy EXC 2044--390685587, Mathematics M\"unster: Dynamics–Geometry–Structure. 
\end{acknowledgments}

\bibliography{main.bib} 

\end{document}

%% file: fdiagrams.tex

\definecolor{jred}{rgb}{0.8,0,0}
\definecolor{jgreen}{rgb}{0,0.7,0}
\definecolor{jblue}{rgb}{0,0,0.8}

\usetikzlibrary{arrows,patterns,shapes,positioning,fit}
\tikzstyle{c} =	[coordinate]
\tikzstyle{vc} = [circle, draw=black, line width=1pt, inner sep=0pt, minimum size=2mm]
\tikzstyle{v} =  [circle, draw=black, line width=.2pt, fill=black, inner sep=0pt, minimum size=1.5mm]
\tikzstyle{vb} =[circle, draw=black, line width=.1pt, fill=jred, inner sep=0pt, minimum size=1mm]
\tikzstyle{vh} =[circle, draw=black, line width=.1pt, fill=jblue, inner sep=0pt, minimum size=1.5mm]
\tikzstyle{vs} =[coordinate]
\tikzstyle{e} =	[draw=jred,line width=1pt]
\tikzstyle{eb}= [draw=jgreen,line width=1pt]
\tikzstyle{es}= [dashed, line width=1pt]
\tikzstyle{ev}= [draw=jgreen,line width=1pt]
\tikzstyle{f} = [line width=0.01pt,dotted,fill=blue, fill opacity=.1]
\tikzstyle{cs}= [ellipse, draw=black, line width=.1pt, fill=jred, inner sep=1pt, minimum size=2mm]

\newcommand{\vertexexstranded}{
\begin{tikzpicture}[x=6ex,y=6ex,baseline={([yshift=-.59ex]current bounding box.center)}] 
    \node [vs]	(12)	at (1,.1)	{};
    \node [vs]	(13)	at (1,0)	{};
    \node [vs]	(14)	at (1,-.1)	{};    
    \node [vs]	(21)	at (.1,1)	{};
    \node [vs]	(24)	at (0,1)	{};
    \node [vs]	(23)	at (-.1,1)	{};    
    \node [vs]	(32)	at (-1,.1)	{};
    \node [vs]	(31)	at (-1,0)	{};
    \node [vs]	(34)	at (-1,-.1)	{};    
    \node [vs]	(41)	at (.1,-1)	{};
    \node [vs]	(42)	at (0,-1)	{};
    \node [vs]	(43)	at (-.1,-1)	{};
    \node [vc, fit=(13) (31)] {};
    \path
    \foreach \i/\j in {14/41,21/12,32/23,43/34}{
    (\i) edge [ev,bend right=40] node [vh] {} (\j)
    }
    \foreach \i/\j in {13/42, 31/24}{
    (\i) edge [ev,bend right=60] node [vh] {} (\j)
    };
    \node [cs,fit=(12) (14)] {};
    \node [cs,fit=(21) (23)] {};
    \node [cs,fit=(32) (34)] {};
    \node [cs,fit=(41) (43)] {};
\end{tikzpicture}
}

\newcommand{\vertexexgraph}{
\begin{tikzpicture}[x=6ex,y=6ex,baseline={([yshift=-.59ex]current bounding box.center)}]    
\scriptsize{
    \node [vb, label=right:${1}$]	(2)	at (0,1)	{};
    \node [vb, label=below:${2}$]	(3)	at (-1,0)	{};
    \node [vb, label=left:${3}$]	(4)	at (0,-1)	{};
    \node [vb,label=above:${4}$]	(1)	at (1,0)	{};
    \node [v, fill=black, minimum size =3mm]   (0) at (0,0)    {};
    \path
    (1) edge [ev] (2)
    (3) edge [ev] (4)
    \foreach \i in {1,2,3,4}{
    (\i) edge [e] (0)
    }
    \foreach \i/\j in {1/4,4/1,2/3,3/2}{
    (\i) edge [ev,bend right=20] node [c] {} (\j)
    };
}
\end{tikzpicture}
}

\newcommand{\vertexexeom}[3]{
\begin{tikzpicture}[x=4ex,y=4ex,baseline={([yshift=-.59ex]current bounding box.center)}]    
\scriptsize{
    \node [vb, label=above:${#1}$]	(2)	at (0,1)	{};
    \node [vb, label=below:${#2}$]	(3)	at (-1,0)	{};
    \node [vb, label=below:${#3}$]	(4)	at (0,-1)	{};
    \node [c]   (a) at (.7,1)   {};
    \node [c]   (b) at (.7,-.8) {};
    \node [c]   (c) at (.7,-1.2){};
    \node [v, fill=black]   (0) at (0,0)    {};
    \path
    (a) edge [ev] (2)
    (3) edge [ev] (4)
    (b) edge [ev] (4)
    (c) edge [ev] (4)
    \foreach \i in {2,3,4}{
    (\i) edge [e] (0)
    }
    \foreach \i/\j in {2/3,3/2}{
    (\i) edge [ev,bend right=20] node [c] {} (\j)
    };
}
\end{tikzpicture}
}

\newcommand{\vertexexvg}{
\begin{tikzpicture}[rotate=45,x=1ex,y=1ex,baseline={([yshift=-.59ex]current bounding box.center)}]    
    \node [vb]	(1)	at (1,0)	{};
    \node [vb]	(2)	at (0,1)	{};
    \node [vb]	(3)	at (-1,0)	{};
    \node [vb]	(4)	at (0,-1)	{};
    \path
    (1) edge [ev] (2)
    (3) edge [ev] (4)
    \foreach \i/\j in {1/4,4/1,2/3,3/2}{
    (\i) edge [ev,bend right=40] node [c] {} (\j)
    };
\end{tikzpicture}
}

\newcommand{\maptadpole}{
\begin{tikzpicture}[x=2ex,y=2ex,baseline={([yshift=-.59ex]current bounding box.center)}]    
    \node [v]	(1)	at (0,0) {};
    \node [c]	(a)	at (-1,0){};
    \node [c]	(b) at (1,0) {};
    \node [c]	(c1)at (0,1) {};
    \path
    (1) edge [e] (a)
    (1) edge [e] (b)
    (1) edge [e, bend right=90] (c1)
    (1) edge [e, bend left=90] (c1);
\end{tikzpicture}
}

\newcommand{\maptadpoledown}{
\begin{tikzpicture}[x=2ex,y=2ex,baseline={([yshift=-.59ex]current bounding box.center)}]    
    \node [v]	(1)	at (0,0) {};
    \node [c]	(a)	at (-1,0){};
    \node [c]	(b) at (1,0) {};
    \node [c]	(c1)at (0,-1) {};
    \path
    (1) edge [e] (a)
    (1) edge [e] (b)
    (1) edge [e, bend right=90] (c1)
    (1) edge [e, bend left=90] (c1);
\end{tikzpicture}
}

\newcommand{\mapsunrise}{
\begin{tikzpicture}[x=2ex,y=2ex,baseline={([yshift=-.59ex]current bounding box.center)}]    
\node [v]	(1)	at (0,0) {};
\node [v]	(2)	at (2,0) {};
\node [c]	(a)	at (-1,0){};
\node [c]	(b) at (3,0) {};
\path
(1) edge [e] (a)
(2) edge [e] (b)
(1) edge [e] (2)
(1) edge [e, bend right=90] (2)
(1) edge [e, bend left=90] (2);
\end{tikzpicture}
}

\newcommand{\mapex}{
\begin{tikzpicture}[x=4ex,y=4ex,baseline={([yshift=-.59ex]current bounding box.center)}]    
    \node [v, label=30:1]	(1)	at (-1.5,0) {};
    \node [v, label=150:2, label=-90:3, label=90:4]	(234)	at (0,0) {};
    \node [v, label=-90:5, label=90:6, label=30:7]	(789)at (1.5,0) {};
    \node [c]	(9)	at (2.25,0) {};
    \path
    (1) edge [e] (234)
    (234) edge [e, bend right=80] (789)
    (234) edge [e, bend left=80] (789)
    (789) edge [e] (9);
\end{tikzpicture}
}

\newcommand{\mapexpq}{
\begin{tikzpicture}[x=3ex,y=3ex,baseline={([yshift=-.59ex]current bounding box.center)}]    
    \node [v]	(1)	at (-1.5,0) {};
    \node [v, label=0:$q$]	(234)	at (0,0) {};
    \node [v, label=30:$p$]	(789)   at (1.5,0) {};
    \node [c]	(9)	at (2.25,0) {};
    \path
    (1) edge [e] (234)
    (234) edge [e, bend right=80] (789)
    (234) edge [e, bend left=80] (789)
    (789) edge [e] (9);
\end{tikzpicture}
}

\newcommand{\mapexstranded}{
\begin{tikzpicture}[x=4ex,y=4ex,baseline={([yshift=-.59ex]current bounding box.center)}]
    \node [vs,label=80:1]	(1a)	at (-2,.05)	{};
    \node [vs]				(1b)	at (-2,-.05)	{};
    \node [c]				(1c)	at (-3,0)	{};
    \node [c]				(1d)	at (-2.5,0)	{};	
    \node [vc, fit=(1a) (1c)] 	{};
    \node [cs,fit=(1a) (1b)] 	{};    
    \node [vs, label=30:2]	(24)	at (-1,.05)	{};
    \node [vs]				(23)	at (-1,-.05)	{};
    \node [c]				(2c)	at (1,-.05)	{};
    \begin{scope}[rotate=120]  
    \node [vs, label=below:3]	(32)	at (-1,.05)	{};
    \node [vs]				(34)	at (-1,-.05)	{};
    \end{scope}
    \begin{scope}[rotate=-120]  
    \node [vs]				(43)	at (-1,.05)	{};
    \node [vs,label=above:4]	(42)	at (-1,-.05)	{};
    \end{scope}
    \node [vc, fit=(23) (2c)] {};
    \node [cs,fit=(24) (23)] {};
    \node [cs,fit=(32) (34)] {};
    \node [cs,fit=(42) (43)] {};
    \path
    \foreach \i/\j in {24/42,32/23,43/34}{
    (\i) edge [ev,bend right=30]  (\j)
    };
    \begin{scope}[xshift=2cm]
    \node [vs, label=0:7]   (89)	at (1,.05)	{};
    \node [vs]				(87)	at (1,-.05)	{};
    \node [c]				(8c)	at (-1,-.05){};
    \begin{scope}[rotate=120]  
    \node [vs, label=80:6]  (97)	at (1,.05)	{};
    \node [vs]				(98)	at (1,-.05)	{};
    \end{scope}
    \begin{scope}[rotate=-120]  
    \node [vs]				(78)	at (1,.05)	{};
    \node [vs,label=-80:5]  (79)	at (1,-.05)	{};
    \end{scope}
    \node [vc, fit=(89) (8c)] {};
    \node [cs,fit=(98) (97)] {};
    \node [cs,fit=(79) (78)] {};
    \path
    \foreach \i/\j in {89/98,97/79,78/87}{
    (\i) edge [ev,bend left=30]  (\j)
    };
    \node [cs,fit=(89) (87)] {};
    \end{scope}    
    \path
    (1a) edge [ev, bend right=70] (1d)
    (1b) edge [ev, bend left=75]	 (1d)
    (1a) edge [ev] (24)
    (1b) edge [ev] (23)
    (42) edge [ev, bend left=60] (98)
    (43) edge [ev, bend left=60] (97)
    (34) edge [ev, bend right=60] (79)
    (32) edge [ev, bend right=60] (78);
\end{tikzpicture}
}

\newcommand{\mapexvg}{
\begin{tikzpicture}[x=4ex,y=4ex,baseline={([yshift=-.59ex]current bounding box.center)}]        
    \node [v]	(v1)	at (-2.5,0) {};
    \node [c]	(c1)	at (-3,0) 	{};
    \node [v]	(234)	at (0,0) {};
    \node [v]	(789)at (2.6,0) {};
    \node [vb, label=90:1]	(1)	at (-2,0)	{};
    \node [vb, label=90:2]	(2)	at (-1,0)	{};
    \node [vb, label=-90:3]	(3)	at (.6,-1)	{};
    \node [vb, label=90:4]	(4)	at (.6,1)	{};
    \node [vb, label=90:7]	(9)	at (3.6,0)	{};
    \node [vb, label=90:6]	(8)	at (2,1)	{};
    \node [vb, label=-90:5]	(7)	at (2,-1)	{};
    \path
    (v1) edge [e] (1)
    \foreach \i in {2,3,4}{
    (\i) edge [e] (234)
    }
    \foreach \i in {7,8,9}{
    (\i) edge [e] (789)
    }
    (1) edge [ev, bend right=90] (c1)
    (1) edge [ev, bend left=90] (c1)
    \foreach \i/\j in {2/3,3/4,4/2,7/8,8/9,9/7}{
    (\i) edge [ev] (\j)
    }
    \foreach \i/\j in {1/2,3/7,4/8}{
    (\i) edge [es] (\j)
    };
\end{tikzpicture}
}


\newcommand{\rtedge}{
\begin{tikzpicture}[x=2ex,y=2ex,baseline={([yshift=-.59ex]current bounding box.center)}] 
\node [v]	(v)	at (0,0) 	{};
\node [vb]	(1)	at (-1,0)	{};
\node [vb]	(2)	at (1,0)	{};
\path
\foreach \i in {1,2}{
    (\i) edge [e] (v)
    }
(1)	edge [ev,bend left=35]	(2)
(1)	edge [ev,bend right=35]	(2);
\end{tikzpicture}
}

\newcommand{\edgevgb}{
\begin{tikzpicture}[x=2ex,y=2ex,baseline={([yshift=-.59ex]current bounding box.center)}]
\node [vb]	(1)	at (0,1)	{};
\node [vb]	(2)	at (0,-1)	{};
\path
(1)	edge [ev,bend left=30]	(2)
(1)	edge [ev,bend right=30]	(2);
\end{tikzpicture}
}

\newcommand{\edgevg}{
\begin{tikzpicture}[x=2ex,y=2ex,baseline={([yshift=-.59ex]current bounding box.center)}]
\node [vb]	(1)	at (1,0)	{};
\node [vb]	(2)	at (-1,0)	{};
\path
(1)	edge [ev,bend left=30]	(2)
(1)	edge [ev,bend right=30]	(2);
\end{tikzpicture}
}

\newcommand{\edgevgr}{
\begin{tikzpicture}[x=2ex,y=2ex,baseline={([yshift=-.59ex]current bounding box.center)}]
\node [vb]	(1)	at (1,0)	{};
\node [vb]	(2)	at (-1,0)	{};
\node at (0,0) {...};
\path
(1)	edge [ev,bend left=50] 	(2)
(1)	edge [ev,bend left=30] 	(2)
(1)	edge [ev,bend right=30] (2)
(1)	edge [ev,bend right=50]	(2);
\end{tikzpicture}
}

\newcommand{\mapthreevertex}{
\begin{tikzpicture}[x=2ex,y=2ex,baseline={([yshift=-.59ex]current bounding box.center)}] 
\node [v]	(v) at (0,0) 	{};
\node [vb]	(1)	at (-1,0)	{};
\node [vb]	(2)	at (.6,-1)	{};
\node [vb]	(3)	at (.6,1)	{};
\path
\foreach \i in {1,2,3}{
    (\i) edge [e] (v)
    }
\foreach \i/\j in {1/2,2/3,3/1}{
    (\i) edge [ev] (\j)
    };
\end{tikzpicture}
}

\newcommand{\onevg}{
\begin{tikzpicture}[x=1.2ex,y=1.2ex,baseline={([yshift=-.59ex]current bounding box.center)}] 
    \node [c]	(c1)	at (-3,0) 	{};
    \node [vb]	(1)	at (-2,0)	{};    
    \path
    (1) edge [ev, bend right=90] (c1)
    (1) edge [ev, bend left=90] (c1);
\end{tikzpicture}
}

\newcommand{\threevg}{
\begin{tikzpicture}[x=.7ex,y=.7ex,baseline={([yshift=-.59ex]current bounding box.center)}] 
\node [vb]	(1)	at (-1,0)	{};
\node [vb]	(2)	at (.6,-1)	{};
\node [vb]	(3)	at (.6,1)	{};
\path
\foreach \i/\j in {1/2,2/3,3/1}{
    (\i) edge [ev] (\j)
    };
\end{tikzpicture}
}

\newcommand{\rtvf}{
\begin{tikzpicture}[x=2ex,y=2ex,baseline={([yshift=-.59ex]current bounding box.center)}]
\scriptsize{
\node [v]	(v)	at (0,0) 	{};
\node [vb]	(1)	at (-1,1)	{};
\node [vb]	(2)	at (-1,-1)	{};
\node [vb]	(3)	at (1,-1)	{};
\node [vb]	(4)	at (1,1)	{};
\path
\foreach \i in {1,2,3,4}{
    (\i) edge [e] (v)
    }
(1) edge [ev]  node [c, label=left:$c$] {} (2)
\foreach \i/\j in {2/3,3/4,4/1}{
    (\i) edge [ev] (\j)
    };
}
\end{tikzpicture}
}

\newcommand{\rtvfg}{
\begin{tikzpicture}[x=.7ex,y=.7ex,baseline={([yshift=-.59ex]current bounding box.center)}]
\node [vb]	(1)	at (-1,1)	{};
\node [vb]	(2)	at (-1,-1)	{};
\node [vb]	(3)	at (1,-1)	{};
\node [vb]	(4)	at (1,1)	{};
\path
\foreach \i/\j in {1/2,2/3,3/4,4/1}{
    (\i) edge [ev] (\j)
    };
\end{tikzpicture}
}

\newcommand{\rttadpole}{
\begin{tikzpicture}[x=2ex,y=2ex,baseline={([yshift=-1.2ex]current bounding box.center)}]
\node [v]	(v)	at (0,0) 	{};
\node [vb]	(1)	at (-1,1)	{};
\node [vb]	(2)	at (-1,-1)	{};
\node [vb]	(3)	at (1,-1)	{};
\node [vb]	(4)	at (1,1)	{};
\path
\foreach \i in {1,2,3,4}{
    (\i) edge [e] (v)
    }
\foreach \i/\j in {1/2,2/3,3/4,4/1}{
    (\i) edge [ev] (\j)
    }
(1)	edge [es,bend left=60]	(4);
\end{tikzpicture}
}

\newcommand{\rttadpolec}{
\begin{tikzpicture}[x=2ex,y=2ex,baseline={([yshift=-1.2ex]current bounding box.center)}]
\scriptsize{
\node [v]	(v)	at (0,0) 	{};
\node [vb]	(1)	at (-1,1)	{};
\node [vb]	(2)	at (-1,-1)	{};
\node [vb]	(3)	at (1,-1)	{};
\node [vb]	(4)	at (1,1)	{};
\path
\foreach \i in {1,2,3,4}{
    (\i) edge [e] (v)
    }
(1) edge [ev]  node [c, label=left:$c$] {} (2)
\foreach \i/\j in {2/3,3/4,4/1}{
    (\i) edge [ev] (\j)
    }
(1)	edge [es,bend left=60]	(4);
}
\end{tikzpicture}
}

\newcommand{\rttadpolepq}{
\begin{tikzpicture}[x=2ex,y=2ex,baseline={([yshift=-1.2ex]current bounding box.center)}]
\scriptsize{
\node [v]	(v)	at (0,0) 	{};
\node [vb]	(1)	at (-1,1)	{};
\node [vb]	(2)	at (-1,-1)	{};
\node [vb]	(3)	at (1,-1)	{};
\node [vb]	(4)	at (1,1)	{};
\path
\foreach \i in {1,2,3,4}{
    (\i) edge [e] (v)
    }
(1) edge [ev]  node [c, label=left:$p_1$] {} (2)
(1) edge [ev]  node [c, label=above:$q_2$]{} (4)
\foreach \i/\j in {2/3,3/4,4/1}{
    (\i) edge [ev] (\j)
    };
\draw [es] (1) arc [start angle=180, end angle=0, radius=1];
}
\end{tikzpicture}
}

\newcommand{\rttadpoleb}{
\begin{tikzpicture}[x=2ex,y=2ex,baseline={([yshift=-.59ex]current bounding box.center)}]
\scriptsize{
\node [v]	(v)	at (0,0) 	{};
\node [vb]	(1)	at (-1,1)	{};
\node [vb]	(2)	at (-1,-1)	{};
\node [vb]	(3)	at (1,-1)	{};
\node [vb]	(4)	at (1,1)	{};
\path
\foreach \i in {1,2,3,4}{
    (\i) edge [e] (v)
    }
(1) edge [ev]  node [c, label=above:$c$] {} (4)
\foreach \i/\j in {1/2,2/3,3/4}{
    (\i) edge [ev] (\j)
    }
(2)	edge [es,bend right=60]	(3);
}
\end{tikzpicture}
}

\newcommand{\rttadpolebpq}{
\begin{tikzpicture}[x=2ex,y=2ex,baseline={([yshift=.49ex]current bounding box.center)}]
\scriptsize{
\node [v]	(v)	at (0,0) 	{};
\node [vb]	(1)	at (-1,1)	{};
\node [vb]	(2)	at (-1,-1)	{};
\node [vb]	(3)	at (1,-1)	{};
\node [vb]	(4)	at (1,1)	{};
\path
\foreach \i in {1,2,3,4}{
    (\i) edge [e] (v)
    }
(1) edge [ev]  node [c, label=left:$p_2$] {} (2)
(2) edge [ev]  node [c, label=below:$q_1$] {} (3)
\foreach \i/\j in {3/4,4/1}{
    (\i) edge [ev] (\j)
    };
\draw [es] (2) arc [start angle=180, end angle=360, radius=1];
}
\end{tikzpicture}
}

\newcommand{\rtsunrisec}{
\begin{tikzpicture}[x=2ex,y=2ex,baseline={([yshift=-.6ex]current bounding box.center)}]
\scriptsize{
\begin{scope}[rotate=-45]
\node [v]	(v1)at (0,0) 	{};
\node [vb]	(1)	at (-1,1)	{};
\node [vb]	(2)	at (-1,-1)	{};
\node [vb]	(3)	at (1,-1)	{};
\node [vb]	(4)	at (1,1)	{};
\end{scope}
\begin{scope}[xshift=1.5cm,rotate=-45]
\node [v]	(v2)at (0,0) 	{};
\node [vb]	(5)	at (-1,1)	{};
\node [vb]	(6)	at (-1,-1)	{};
\node [vb]	(7)	at (1,-1)	{};
\node [vb]	(8)	at (1,1)	{};
\end{scope}
\path
\foreach \i in {1,2,3,4}{
    (\i) edge [e] (v1)
    }
\foreach \i in {5,6,7,8}{
    (\i) edge [e] (v2)
    }
(1) edge [ev]  node [c, label=135:$c$] {} (2)
(8) edge [ev]  node [c, label=45:$c$] {} (5)
\foreach \i/\j in {2/3,3/4,4/1,5/6,6/7,7/8}{
    (\i) edge [ev] (\j)
    }
\foreach \i/\j in {1/5,3/7,4/6}{
    (\i) edge [es] (\j)
    };
}
\end{tikzpicture}
}

\newcommand{\rtsunrisepq}{
\begin{tikzpicture}[x=2ex,y=2ex,baseline={([yshift=-.6ex]current bounding box.center)}]
\scriptsize{
\begin{scope}[rotate=-45]
\node [v]	(v1)at (0,0) 	{};
\node [vb]	(1)	at (-1,1)	{};
\node [vb]	(2)	at (-1,-1)	{};
\node [vb]	(3)	at (1,-1)	{};
\node [vb]	(4)	at (1,1)	{};
\end{scope}
\begin{scope}[xshift=1.5cm,rotate=-45]
\node [v]	(v2)at (0,0) 	{};
\node [vb]	(5)	at (-1,1)	{};
\node [vb]	(6)	at (-1,-1)	{};
\node [vb]	(7)	at (1,-1)	{};
\node [vb]	(8)	at (1,1)	{};
\end{scope}
\path
\foreach \i in {1,2,3,4}{
    (\i) edge [e] (v1)
    }
\foreach \i in {5,6,7,8}{
    (\i) edge [e] (v2)
    }
(1) edge [ev]  node [c, label=135:$p_1$] {} (2)
(2) edge [ev]  node [c, label=225:$p_2$] {} (3)
(3) edge [ev]  node [c, label=right:$q_1$] {} (4)
(4) edge [ev]  node [c, label=right:$q_2$] {} (1)
\foreach \i/\j in {5/6,6/7,7/8,8/5}{
    (\i) edge [ev] (\j)
    }
\foreach \i/\j in {1/5,3/7,4/6}{
    (\i) edge [es] (\j)
    };
}
\end{tikzpicture}
}

\newcommand{\rtsunrisesub}[1]{
\begin{tikzpicture}[x=2ex,y=2ex,baseline={([yshift=-.6ex]current bounding box.center)}]
\scriptsize{
\begin{scope}[rotate=-45]
\node [v]	(v1)at (0,0) 	{};
\node [vb]	(1)	at (-1,1)	{};
\node [vb]	(2)	at (-1,-1)	{};
\node [vb]	(3)	at (1,-1)	{};
\node [vb]	(4)	at (1,1)	{};
\end{scope}
\begin{scope}[xshift=1.5cm,rotate=-45]
\node [v]	(v2)at (0,0) 	{};
\node [vb]	(5)	at (-1,1)	{};
\node [vb]	(6)	at (-1,-1)	{};
\node [vb]	(7)	at (1,-1)	{};
\node [vb]	(8)	at (1,1)	{};
\end{scope}
\path
\foreach \i in {1,2,3,4}{
    (\i) edge [e] (v1)
    }
\foreach \i in {5,6,7,8}{
    (\i) edge [e] (v2)
    }
(1) edge [ev]  node [c, label=135:$c$] {} (2)
(8) edge [ev]  node [c, label=45:$c$] {} (5)
\foreach \i/\j in {2/3,3/4,4/1,5/6,6/7,7/8}{
    (\i) edge [ev] (\j)
    }
\foreach \i/\j in {#1}{
    (\i) edge [es] (\j)
    };
}
\end{tikzpicture}
}

\newcommand{\rtfisha}{
\begin{tikzpicture}[x=2ex,y=2ex,baseline={([yshift=-1.2ex]current bounding box.center)}]
\scriptsize{
\begin{scope}[rotate=-90]
\node [v]	(v1)at (0,0) 	{};
\node [vb]	(1)	at (-1,1)	{};
\node [vb]	(2)	at (-1,-1)	{};
\node [vb]	(3)	at (1,-1)	{};
\node [vb]	(4)	at (1,1)	{};
\end{scope}
\begin{scope}[xshift=1.3cm]
\node [v]	(v2)at (0,0) 	{};
\node [vb]	(5)	at (-1,1)	{};
\node [vb]	(6)	at (-1,-1)	{};
\node [vb]	(7)	at (1,-1)	{};
\node [vb]	(8)	at (1,1)	{};
\end{scope}
\path
\foreach \i in {1,2,3,4}{
    (\i) edge [e] (v1)
    }
\foreach \i in {5,6,7,8}{
    (\i) edge [e] (v2)
    }
(1) edge [ev]  node [c, label=above:$c$] {} (2)
(8) edge [ev]  node [c, label=above:$c$] {} (5)
\foreach \i/\j in {2/3,3/4,4/1,5/6,6/7,7/8}{
    (\i) edge [ev] (\j)
    }
\foreach \i/\j in {1/5,4/6}{
    (\i) edge [es] (\j)
    };
}
\end{tikzpicture}
}

\newcommand{\rtfishpq}[3]{
\begin{tikzpicture}[x=2ex,y=2ex,baseline={([yshift=-.59ex]current bounding box.center)}]
\scriptsize{
\begin{scope}[rotate=-90]
\node [v]	(v1)at (0,0) 	{};
\node [vb]	(1)	at (-1,1)	{};
\node [vb]	(2)	at (-1,-1)	{};
\node [vb]	(3)	at (1,-1)	{};
\node [vb]	(4)	at (1,1)	{};
\end{scope}
\begin{scope}[xshift=1.3cm]
\node [v]	(v2)at (0,0) 	{};
\node [vb]	(5)	at (-1,1)	{};
\node [vb]	(6)	at (-1,-1)	{};
\node [vb]	(7)	at (1,-1)	{};
\node [vb]	(8)	at (1,1)	{};
\end{scope}
\path
\foreach \i in {1,2,3,4}{
    (\i) edge [e] (v1)
    }
\foreach \i in {5,6,7,8}{
    (\i) edge [e] (v2)
    }
(1) edge [ev]  node [c, label=above:$#1$] {} (2)
(3) edge [ev]  node [c, label=below:$#3$] {} (4)
(1) edge [ev]  node [c, label=right:$#2$] {} (4)
\foreach \i/\j in {2/3,3/4,4/1,5/6,6/7,7/8,8/5}{
    (\i) edge [ev] (\j)
    }
\foreach \i/\j in {1/5,4/6}{
    (\i) edge [es] (\j)
    };
}
\end{tikzpicture}
}

\newcommand{\rtfishb}{
\begin{tikzpicture}[x=2ex,y=2ex,baseline={([yshift=-.59ex]current bounding box.center)}]
\scriptsize{
\node [v]	(v1)at (0,0) 	{};
\node [vb]	(1)	at (-1,1)	{};
\node [vb]	(2)	at (-1,-1)	{};
\node [vb]	(3)	at (1,-1)	{};
\node [vb]	(4)	at (1,1)	{};
\begin{scope}[xshift=1.3cm]
\node [v]	(v2)at (0,0) 	{};
\node [vb]	(5)	at (-1,1)	{};
\node [vb]	(6)	at (-1,-1)	{};
\node [vb]	(7)	at (1,-1)	{};
\node [vb]	(8)	at (1,1)	{};
\end{scope}
\path
\foreach \i in {1,2,3,4}{
    (\i) edge [e] (v1)
    }
\foreach \i in {5,6,7,8}{
    (\i) edge [e] (v2)
    }
(1) edge [ev]  node [c, label=left:$c$] {} (2)
(5) edge [ev]  node [c, label=left:$c$] {} (6)
\foreach \i/\j in {2/3,3/4,4/1,6/7,7/8,8/5}{
    (\i) edge [ev] (\j)
    }
\foreach \i/\j in {4/5,3/6}{
    (\i) edge [es] (\j)
    };
}
\end{tikzpicture}
}

\newcommand{\mapa}{
\begin{tikzpicture}[x=2ex,y=2ex,baseline={([yshift=-.59ex]current bounding box.center)}] 
\node [v]	(v)	at (0,0) 	{};
\node [vb]	(1)	at (-1,1)	{};
\node [vb]	(2)	at (-1,-1)	{};
\node [vb]	(3)	at (1,-1)	{};
\node [vb]	(4)	at (1,1)	{};
\path
\foreach \i in {1,2,3,4}{
    (\i) edge [e] (v)
    }
\foreach \i/\j in {1/2,3/4}{
   (\i)	edge [ev,bend left=30]	(\j)
   (\i)	edge [ev,bend right=30]	(\j)
   };
\end{tikzpicture}
}

\newcommand{\mapb}{
\begin{tikzpicture}[x=2ex,y=2ex,baseline={([yshift=-.59ex]current bounding box.center)}] 
\node [v]	(v1)at (0,0) 	{};
\node [v]	(v2)at (2.6,0) 	{};
\node [vb]	(1)	at (-1,0)	{};
\node [vb]	(2)	at (.6,-1)	{};
\node [vb]	(3)	at (.6,1)	{};
\node [vb]	(9)	at (3.6,0)	{};
\node [vb]	(8)	at (2,1)	{};
\node [vb]	(7)	at (2,-1)	{};
\path
\foreach \i in {1,2,3}{
    (\i) edge [e] (v1)
    }
\foreach \i in {7,8,9}{
    (\i) edge [e] (v2)
    }
\foreach \i/\j in {1/2,2/3,3/1,7/8,8/9,9/7}{
    (\i) edge [ev] (\j)
    }
\foreach \i/\j in {2/7,3/8}{
    (\i) edge [es] (\j)
    };
\end{tikzpicture}
}

\newcommand{\mapc}{
\begin{tikzpicture}[x=2ex,y=2ex,baseline={([yshift=-.59ex]current bounding box.center)}]
\node [v]	(v1)	at (0,0) 	{};
\node [vb]	(1)	at (-1,1)	{};
\node [vb]	(2)	at (-1,-1)	{};
\node [vb]	(3)	at (1,-1)	{};
\node [vb]	(4)	at (1,1)	{};
\begin{scope}[xshift=7ex]
\node [v]	(v2)	at (0,0) 	{};
\node [vb]	(5)	at (-1,1)	{};
\node [vb]	(6)	at (-1,-1)	{};
\node [vb]	(7)	at (1,-1)	{};
\node [vb]	(8)	at (1,1)	{};
\end{scope}
\path
\foreach \i in {1,2,3,4}{
    (\i) edge [e] (v1)
    }
\foreach \i in {5,6,7,8}{
    (\i) edge [e] (v2)
    }
\foreach \i/\j in {1/2,2/3,3/4,4/1,5/6,6/7,7/8,8/5}{
    (\i) edge [ev] (\j)
    }
\foreach \i/\j in {4/8,7/3}{
    (\i) edge [es, bend left=30] (\j)
    };
\end{tikzpicture}
}

\newcommand{\mapd}{
\begin{tikzpicture}[x=2ex,y=2ex,baseline={([yshift=-.59ex]current bounding box.center)}]
\node [v]	(v1)at (0,0) 	{};
\node [v]	(v2)at (2.4,0) 	{};
\node [vb]	(1)	at (-1,0)	{};
\node [vb]	(2)	at (.6,-1)	{};
\node [vb]	(3)	at (.6,1)	{};
\node [vb]	(4)	at (1.4,0)	{};
\node [vb]	(5)	at (3,-1)	{};
\node [vb]	(6)	at (3,1)	{};
\path
\foreach \i in {1,2,3}{
    (\i) edge [e] (v1)
    }
\foreach \i in {4,5,6}{
    (\i) edge [e] (v2)
    }
\foreach \i/\j in {1/2,2/3,3/1,4/5,5/6,6/4}{
    (\i) edge [ev] (\j)
    }
\foreach \i/\j in {2/5,3/6}{
    (\i) edge [es] (\j)
    };
\end{tikzpicture}
}

\newcommand{\maptorus}{
\begin{tikzpicture}[x=2ex,y=2ex,baseline={([yshift=-.59ex]current bounding box.center)}]
\node [v]	(v3)	at (0,0) 	{};
\node [v]	(v4)	at (2,0) 	{};
\node [vb]	(11)	at (-1,0)	{};
\node [vb]	(12)	at (.6,-1)	{};
\node [vb]	(13)	at (.6,1)	{};
\node [vb]	(9)	at (3,0)	{};
\node [vb]	(8)	at (1.4,1)	{};
\node [vb]	(7)	at (1.4,-1)	{};
\begin{scope}[yshift=5ex]
\node [v]	(v1)	at (0,0) 	{};
\node [v]	(v2)	at (2.4,0) 	{};
\node [vb]	(1)	at (-1,0)	{};
\node [vb]	(2)	at (.6,-1)	{};
\node [vb]	(3)	at (.6,1)	{};
\node [vb]	(4)	at (1.4,0)	{};
\node [vb]	(5)	at (3,-1)	{};
\node [vb]	(6)	at (3,1)	{};
\end{scope}
\path
\foreach \i/\j in {1/v1,2/v1,3/v1,4/v2,5/v2,6/v2,11/v3,12/v3,13/v3,7/v4,8/v4,9/v4}{
    (\i) edge [e] (\j)
    }
\foreach \i/\j in {1/2,2/3,3/1,4/5,5/6,6/4,7/8,8/9,9/7,11/12,12/13,13/11}{
    (\i) edge [ev] (\j)
    }
\foreach \i/\j in {2/5,3/6,1/11,12/7,13/8}{
    (\i) edge [es] (\j)
    }
(4) edge [es,sloped] (9);
\end{tikzpicture}
}

\newcommand{\mapcon}{
\begin{tikzpicture}[x=2ex,y=2ex,baseline={([yshift=-.59ex]current bounding box.center)}]
\draw [ev] (0,1.8) circle (.5ex); 
\draw [ev] (2,1.8) circle (.5ex); 
\node [v]	(v3)at (0,0) 	{};
\node [v]	(v4)at (2,0) 	{};
\node [vb]	(1)	at (-1,0)	{};
\node [vb]	(2)	at (.6,-1)	{};
\node [vb]	(3)	at (.6,1)	{};
\node [vb]	(9)	at (3,0)	{};
\node [vb]	(8)	at (1.4,1)	{};
\node [vb]	(7)	at (1.4,-1)	{};
\node [v]	(v2)at (1,1.5) 	{};
\node [vb]	(4)	at (0,1.5)	{};
\node [vb]	(5)	at (2,1.5)	{};
\path
\foreach \i/\j in {4/v2,5/v2,1/v3,2/v3,3/v3,7/v4,8/v4,9/v4}{
    (\i) edge [e] (\j)
    }
\foreach \i/\j in {1/2,2/3,3/1,7/8,8/9,9/7}{
    (\i) edge [ev] (\j)
    }
\foreach \i/\j in {2/7,3/8,1/4,9/5}{
    (\i) edge [es] (\j)
    };
\end{tikzpicture}
}


\newcommand{\rhvfc}{
\begin{tikzpicture}[rotate=90, x=2ex,y=2ex,baseline={([yshift=-.59ex]current bounding box.center)}] 
\scriptsize{
\node [v]	(1)	at (0,0)	{};
\node [vb]	(b1)	at (-1,1)	{};
\node [vb]	(w1)	at (-1,-1)	{};
\node [vb]	(b2)	at (1,-1)	{};
\node [vb]	(w2)	at (1,1)	{};
\path
\foreach \i/\j in {1/2}{
   	(w\i) edge [ev]   (b\j)
	(b\i) edge [ev]  node [c, label=left:$c$] {} (w\j)
	}
\foreach \i in {1,2}{
	(w\i)	edge [ev,bend left=30]	(b\i)
	(w\i)	edge [ev,bend right=30]	(b\i)
	}
\foreach \i in {w1,b1,w2,b2}{
   	(\i) edge [e] (1)
	};
}
\end{tikzpicture}
}

\newcommand{\rhvfg}{
\begin{tikzpicture}[rotate=-90,x=.7ex,y=.7ex,baseline={([yshift=-.59ex]current bounding box.center)}] 
\scriptsize{
\node [vb]	(b1)	at (-1,1)	{};
\node [vb]	(w1)	at (-1,-1)	{};
\node [vb]	(b2)	at (1,-1)	{};
\node [vb]	(w2)	at (1,1)	{};
\path
\foreach \i/\j in {1/2}{
   	(w\i) edge [ev]   (b\j)
	(b\i) edge [ev]  node [c, label=right:$c$] {} (w\j)
	}
\foreach \i in {1,2}{
	(w\i)	edge [ev,bend left=30]	(b\i)
	(w\i)	edge [ev,bend right=30]	(b\i)
	};
}
\end{tikzpicture}
}

\newcommand{\rhtadpolea}{
\begin{tikzpicture}[rotate=90, x=2ex,y=2ex,baseline={([yshift=-1.2ex]current bounding box.center)}] 
\scriptsize{
\node [v]	(1)	    at (0,0)	{};
\node [vb]	(b1)	at (-1,1)	{};
\node [vb]	(w1)	at (-1,-1)	{};
\node [vb]	(b2)	at (1,-1)	{};
\node [vb]	(w2)	at (1,1)	{};
\path
\foreach \i/\j in {1/2}{
   	(w\i) edge [ev]   (b\j)
	(b\i) edge [ev]  node [c, label=left:$c$] {} (w\j)
	}
\foreach \i in {1,2}{
	(w\i)	edge [ev,bend left=30]	(b\i)
	(w\i)	edge [ev,bend right=30]	(b\i)
	}
\foreach \i in {w1,b1,w2,b2}{
   	(\i) edge [e] (1)
	}
(w2)	edge [es,bend left=90]	(b2);
}
\end{tikzpicture}
}

\newcommand{\rhtadpoleapq}{
\begin{tikzpicture}[rotate=90, x=2ex,y=2ex,baseline={([yshift=-1.2ex]current bounding box.center)}] 
\scriptsize{
\node [v]	(1)	    at (0,0)	{};
\node [vb]	(b1)	at (-1,1)	{};
\node [vb]	(w1)	at (-1,-1)	{};
\node [vb]	(b2)	at (1,-1)	{};
\node [vb]	(w2)	at (1,1)	{};
\path
\foreach \i/\j in {1/2}{
   	(w\i) edge [ev]   (b\j)
	(b\i) edge [ev]  node [c, label=left:$p_1$] {} (w\j)
	}
\foreach \i in {1,2}{
	(w\i)	edge [ev,bend left=30]	(b\i)
	(w\i)	edge [ev,bend right=30]	(b\i)
	}
\foreach \i in {w1,b1,w2,b2}{
   	(\i) edge [e] (1)
	}
(w2)	edge [es,bend left=90]	(b2);
}
\end{tikzpicture}
}

\newcommand{\rhtadpoleb}{
\begin{tikzpicture}[x=2ex,y=2ex,baseline={([yshift=-.3ex]current bounding box.center)}] 
\scriptsize{
\node [v]	(1)	    at (0,0)	{};
\node [vb]	(b1)	at (-1,1)	{};
\node [vb]	(w1)	at (-1,-1)	{};
\node [vb]	(b2)	at (1,-1)	{};
\node [vb]	(w2)	at (1,1)	{};
\path
\foreach \i/\j in {1/2}{
   	(w\i) edge [ev]  node [c, label=below:$c$] {} (b\j)
	(b\i) edge [ev]   (w\j)
	}
\foreach \i in {1,2}{
	(w\i)	edge [ev,bend left=30]	(b\i)
	(w\i)	edge [ev,bend right=30]	(b\i)
	}
\foreach \i in {w1,b1,w2,b2}{
   	(\i) edge [e] (1)
	}
(b1)	edge [es,bend left=60]	(w2);
}
\end{tikzpicture}
}

\newcommand{\rhtadpolebpq}{
\begin{tikzpicture}[x=2ex,y=2ex,baseline={([yshift=-.3ex]current bounding box.center)}] 
\scriptsize{
\node [v]	(1)	    at (0,0)	{};
\node [vb]	(b1)	at (-1,1)	{};
\node [vb]	(w1)	at (-1,-1)	{};
\node [vb]	(b2)	at (1,-1)	{};
\node [vb]	(w2)	at (1,1)	{};
\path
\foreach \i/\j in {1/2}{
   	(w\i) edge [ev]  (b\j)
	(b\i) edge [ev]  node [c, label=above:$q_1$] {} (w\j)
	}
(w1)	edge [ev,bend left=30] node [c, label=left:{$p_2,p_3$}] {}	(b1)
(w1)	edge [ev,bend right=30]	(b1)
\foreach \i in {2}{
	(w\i)	edge [ev,bend left=30]	(b\i)
	(w\i)	edge [ev,bend right=30]	(b\i)
	}
\foreach \i in {w1,b1,w2,b2}{
   	(\i) edge [e] (1)
	};
\draw [es] (b1) arc [start angle=180, end angle=0, radius=1];
}
\end{tikzpicture}
}

\newcommand{\rhsunrise}[5]{
\begin{tikzpicture}[x=2ex,y=2ex,baseline={([yshift=-.6ex]current bounding box.center)}]
\scriptsize{
\begin{scope}[rotate=-45]
\node [v]	(v1)at (0,0) 	{};
\node [vb]	(1)	at (-1,1)	{};
\node [vb]	(2)	at (-1,-1)	{};
\node [vb]	(3)	at (1,-1)	{};
\node [vb]	(4)	at (1,1)	{};
\end{scope}
\begin{scope}[xshift=1.5cm,rotate=-45]
\node [v]	(v2)at (0,0) 	{};
\node [vb]	(5)	at (-1,1)	{};
\node [vb]	(6)	at (-1,-1)	{};
\node [vb]	(7)	at (1,-1)	{};
\node [vb]	(8)	at (1,1)	{};
\end{scope}
\path
\foreach \i in {1,2,3,4}{
    (\i) edge [e] (v1)
    }
\foreach \i in {5,6,7,8}{
    (\i) edge [e] (v2)
    }
(1) edge [ev]  node [c, label=135:$#1$] {} (2)
(2) edge [ev, bend left=30]  node [c, label=225:$#2$] {} (3)
(2) edge [ev, bend right=30] (3)
(4) edge [ev, bend right=30]  node [c, label=right:$#3$] {} (1)
(4) edge [ev, bend left=30]   (1)
(3) edge [ev]  node [c, label=right:$#4$] {} (4)
(5) edge [ev]  node [c, label=45:$#5$] {} (8)
\foreach \i/\j in {6/7,8/5}{
    (\i) edge [ev] (\j)
    }
\foreach \i\j in {5/6,7/8}{
	(\i)	edge [ev,bend left=30]	(\j)
	(\i)	edge [ev,bend right=30]	(\j)
	}
\foreach \i/\j in {1/5,3/7,4/6}{
    (\i) edge [es] (\j)
    };
}
\end{tikzpicture}
}

\newcommand{\rhfish}[4]{
\begin{tikzpicture}[x=2ex,y=2ex,baseline={([yshift=-.59ex]current bounding box.center)}]
\scriptsize{
\begin{scope}[rotate=-90]
\node [v]	(v1)at (0,0) 	{};
\node [vb]	(1)	at (-1,1)	{};
\node [vb]	(2)	at (-1,-1)	{};
\node [vb]	(3)	at (1,-1)	{};
\node [vb]	(4)	at (1,1)	{};
\end{scope}
\begin{scope}[xshift=1.8cm]
\node [v]	(v2)at (0,0) 	{};
\node [vb]	(5)	at (-1,1)	{};
\node [vb]	(6)	at (-1,-1)	{};
\node [vb]	(7)	at (1,-1)	{};
\node [vb]	(8)	at (1,1)	{};
\end{scope}
\path
\foreach \i in {1,2,3,4}{
    (\i) edge [e] (v1)
    }
\foreach \i in {5,6,7,8}{
    (\i) edge [e] (v2)
    }
(1) edge [ev]  node [c, label=above:$#1$] {} (2)
(1) edge [ev,bend left=30]  node [c, label=right:$#2$] {} (4)
(1) edge [ev,bend right=30] (4)
(3) edge [ev]  node [c, label=below:$#3$] {} (4)
(5) edge [ev]  node [c, label=above:$#4$] {} (8)
\foreach \i/\j in {3/4,6/7,8/5}{
    (\i) edge [ev] (\j)
    }
\foreach \i\j in {2/3,5/6,7/8}{
	(\i)	edge [ev,bend left=30]	(\j)
	(\i)	edge [ev,bend right=30]	(\j)
	}
\foreach \i/\j in {1/5,4/6}{
    (\i) edge [es] (\j)
    };
}
\end{tikzpicture}
}


\newcommand{\cvtvg}{
\begin{tikzpicture}[x=1ex,y=1ex,baseline={([yshift=-.59ex]current bounding box.center)}] 
\node [vb]	(1)	at (0,1)	{};
\node [vb]	(2)	at (0,-1)	{};
\path
\foreach \i/\j in {1/2,2/1}{
   (\i)	edge [ev,bend left=15]	(\j)
   (\i)	edge [ev,bend left=40]	(\j)
   };
\end{tikzpicture}
}

\newcommand{\cvt}{
\begin{tikzpicture}[x=2ex,y=2ex,baseline={([yshift=-.59ex]current bounding box.center)}] 
\node [v]	(v)	at (0,0)	{};
\node [vb]	(1)	at (1,0)	{};
\node [vb]	(2)	at (-1,0)	{};
\path
\foreach \i in {1,2}{
    (\i) edge [e] (v)
    }
\foreach \i/\j in {1/2,2/1}{
   (\i)	edge [ev,bend left=30]	(\j)
   (\i)	edge [ev,bend left=50]	(\j)
   };
\end{tikzpicture}
}

\newcommand{\cvfvg}{
\begin{tikzpicture}[x=1ex,y=1ex,baseline={([yshift=-.59ex]current bounding box.center)}] 
\scriptsize{
\node [vb, label=above:]	(b1)	at (-1,1)	{};
\node [vb, label=below:]	(w1)	at (-1,-1)	{};
\node [vb, label=below:]	(b2)	at (1,-1)	{};
\node [vb, label=above:]	(w2)	at (1,1)	{};
\path
\foreach \i/\j in {1/2}{
   	(w\i) edge [ev]  node [c, label=below:$\ell$] {} (b\j)
	(b\i) edge [ev]   (w\j)
	}
\foreach \i in {1,2}{
	(w\i)   edge [ev]				(b\i)
	(w\i)	edge [ev,bend left=30]	(b\i)
	(w\i)	edge [ev,bend right=30]	(b\i)
	};
}
\end{tikzpicture}
}

\newcommand{\cvfnvg}{
\begin{tikzpicture}[x=1ex,y=1ex,baseline={([yshift=-.59ex]current bounding box.center)}] 
\scriptsize{
\node [vb, label=above:]	(b1)	at (-1,1)	{};
\node [vb, label=below:]	(w1)	at (-1,-1)	{};
\node [vb, label=below:]	(b2)	at (1,-1)	{};
\node [vb, label=above:]	(w2)	at (1,1)	{};
\path
\foreach \i/\j in {1/2,2/1}{
   	(w\i) edge [ev,bend left=30]  (b\j)
   	(w\i) edge [ev,bend right=30] (b\j)
	}
\foreach \i in {1,2}{
	(w\i)	edge [ev,bend left=30]  (b\i)
	(w\i)	edge [ev,bend right=30]	(b\i)
	}

\foreach \i in {w1,b1,w2,b2}{
	};
}
\end{tikzpicture}
}

\newcommand{\cvf}{
\begin{tikzpicture}[x=2ex,y=2ex,baseline={([yshift=-.59ex]current bounding box.center)}] 
\scriptsize{
\node [v]	(1)	at (0,0)	{};
\node [vb, label=above:]	(b1)	at (-1,1)	{};
\node [vb, label=below:]	(w1)	at (-1,-1)	{};
\node [vb, label=below:]	(b2)	at (1,-1)	{};
\node [vb, label=above:]	(w2)	at (1,1)	{};
\path
\foreach \i/\j in {1/2}{
   	(w\i) edge [ev]   (b\j)
	(b\i) edge [ev]  node [c, label=above:$\ell$] {} (w\j)
	}
\foreach \i in {1,2}{
	(w\i)   edge 	[ev]				(b\i)
	(w\i)	edge [ev,bend left=30]	(b\i)
	(w\i)	edge [ev,bend right=30]	(b\i)
	}
\foreach \i in {w1,b1,w2,b2}{
   	(\i) edge [e] (1)
	};
}
\end{tikzpicture}
}

\newcommand{\cvfl}{
\begin{tikzpicture}[x=2ex,y=2ex,baseline={([yshift=-.59ex]current bounding box.center)}] 
\scriptsize{
\node [v]	(1)	at (0,0)	{};
\node [vb, label=above:1]	(b1)	at (-1,1)	{};
\node [vb, label=below:2]	(w1)	at (-1,-1)	{};
\node [vb, label=below:7]	(b2)	at (1,-1)	{};
\node [vb, label=above:8]	(w2)	at (1,1)	{};
\path
\foreach \i/\j in {1/2}{
   	(w\i) edge [ev]   (b\j)
	(b\i) edge [ev]  node [c, label=above:$c$] {} (w\j)
	}
\foreach \i in {1,2}{
	(w\i)   edge 	[ev]				(b\i)
	(w\i)	edge [ev,bend left=30]	(b\i)
	(w\i)	edge [ev,bend right=30]	(b\i)
	}
\foreach \i in {w1,b1,w2,b2}{
   	(\i) edge [e] (1)
	};
}
\end{tikzpicture}
}

\newcommand{\cvft}{
\begin{tikzpicture}[x=2ex,y=2ex,baseline={([yshift=-.59ex]current bounding box.center)}] 
\node [v]	(v)	at (0,0) 	{};
\node [vb]	(1)	at (-1,1)	{};
\node [vb]	(2)	at (-1,-1)	{};
\node [vb]	(3)	at (1,-1)	{};
\node [vb]	(4)	at (1,1)	{};
\path
\foreach \i in {1,2,3,4}{
    (\i) edge [e] (v)
    }
\foreach \i/\j in {1/2,2/1,3/4,4/3}{
   (\i)	edge [ev,bend left=15]	(\j)
   (\i)	edge [ev,bend left=40]	(\j)
   };
\end{tikzpicture}
}

\newcommand{\cvfn}{
\begin{tikzpicture}[x=2ex,y=2ex,baseline={([yshift=-.59ex]current bounding box.center)}] 
\scriptsize{
\node [v]	(1)	at (0,0)	{};
\node [vb, label=above:]	(b1)	at (-1,1)	{};
\node [vb, label=below:]	(w1)	at (-1,-1)	{};
\node [vb, label=below:]	(b2)	at (1,-1)	{};
\node [vb, label=above:]	(w2)	at (1,1)	{};
\path
\foreach \i/\j in {1/2,2/1}{
   	(w\i) edge [ev,bend left=30]  (b\j)
   	(w\i) edge [ev,bend right=30] (b\j)
	}
\foreach \i in {1,2}{
	(w\i)	edge [ev,bend left=30]  (b\i)
	(w\i)	edge [ev,bend right=30]	(b\i)
	}
\foreach \i in {w1,b1,w2,b2}{
   	(\i) edge [e] (1)
	};
}
\end{tikzpicture}
}

\newcommand{\cvfsvg}{
\begin{tikzpicture}[x=1ex,y=1ex,baseline={([yshift=-.59ex]current bounding box.center)}] 
\scriptsize{
\foreach \i in {72,144,216,288,360}{
\begin{scope}[rotate=\i]
\node [vb]	(\i)	at (0,1.6)	{};
\end{scope}
}
\path
\foreach \i/\j in {72/144,72/216,72/288,72/360,144/216,144/288,144/360,216/288,216/360,288/360}{
   	(\i) edge [ev]  (\j)
	};
}
\end{tikzpicture}
}

\newcommand{\cvfs}{
\begin{tikzpicture}[x=2ex,y=2ex,baseline={([yshift=-.59ex]current bounding box.center)}] 
\scriptsize{
\node [v]	(0)	at (0,0)	{};
\foreach \i in {72, 144, 216, 288, 360}{
\begin{scope}[rotate=\i]
\node [vb]	(\i)	at (0, 1.6)	{};
\end{scope}
}
\path
\foreach \i/\j in {72/144,72/216,72/288,72/360,144/216,144/288,144/360,216/288,216/360,288/360}{
   	(\i) edge [ev]  (\j)
    }
\foreach \i in {72,144,216,288,360}{
  	(\i) edge [e] (0)
	};
}
\end{tikzpicture}
}

\newcommand{\cvftl}{
\begin{tikzpicture}[x=2ex,y=2ex,baseline={([yshift=-.59ex]current bounding box.center)}] 
\scriptsize{
\node [v]	(v)	at (0,0) 	{};
\node [vb, label=above:1]	(1)	at (-1,1)	{};
\node [vb, label=below:2]	(2)	at (-1,-1)	{};
\node [vb, label=below:7]	(3)	at (1,-1)	{};
\node [vb, label=above:8]	(4)	at (1,1)	{};
\path
\foreach \i in {1,2,3,4}{
    (\i) edge [e] (v)
    }
\foreach \i/\j in {1/2,2/1,3/4,4/3}{
   (\i)	edge [ev,bend left=15]	(\j)
   (\i)	edge [ev,bend left=40]	(\j)
   };
}
\end{tikzpicture}
}

\newcommand{\cfishl}{
\begin{tikzpicture}[x=2ex,y=2ex,baseline={([yshift=-.59ex]current bounding box.center)}] 
\scriptsize{
\node [v]	(1)	at (0,0)	{};
\node [v]	(2)	at (4,0)	{};
\node [vb, label=above:1]	(b1)	at (-1,1)	{};
\node [vb, label=below:2]	(w1)	at (-1,-1)	{};
\node [vb, label=below:3]	(b2)	at (1,-1)	{};
\node [vb, label=above:4]	(w2)	at (1,1)	{};
\node [vb, label=above:5]	(b3)	at (3,1)	{};
\node [vb, label=below:6]	(w3)	at (3,-1)	{};
\node [vb, label=below:7]	(b4)	at (5,-1)	{};
\node [vb, label=above:8]	(w4)	at (5,1)	{};
\path
\foreach \i/\j in {1/2}{
   	(w\i) edge [ev]   (b\j)
	(b\i) edge [ev]  node [c, label=above:$c_1$] {} (w\j)
	}
\foreach \i/\j in {3/4}{
   	(w\i) edge [ev]   (b\j)
	(b\i) edge [ev]  node [c, label=above:$c_2$] {} (w\j)
	}
\foreach \i in {1,2,3,4}{
	(w\i) edge 	[ev]				(b\i)
	(w\i)	edge [ev,bend left=30]	(b\i)
	(w\i)	edge [ev,bend right=30]	(b\i)
	}
\foreach \i in {w1,b1,w2,b2}{
   	(\i) edge [e] (1)
	}
\foreach \i in {w3,b3,w4,b4}{
   	(\i) edge [e] (2)
	}	
\foreach \i/\j in {w2/b3,w3/b2}{
   	(\i) edge [es] (\j)
	};
}
\end{tikzpicture}
}

\newcommand{\cfishlc}{
\begin{tikzpicture}[x=2ex,y=2ex,baseline={([yshift=-.59ex]current bounding box.center)}] 
\scriptsize{
\node [v]	(1)	at (0,0)	{};
\node [v]	(2)	at (4,0)	{};
\node [vb, label=above:1]	(b1)	at (-1,1)	{};
\node [vb, label=below:2]	(w1)	at (-1,-1)	{};
\node [vb, label=below:3]	(b2)	at (1,-1)	{};
\node [vb, label=above:4]	(w2)	at (1,1)	{};
\node [vb, label=above:5]	(b3)	at (3,1)	{};
\node [vb, label=below:6]	(w3)	at (3,-1)	{};
\node [vb, label=below:7]	(b4)	at (5,-1)	{};
\node [vb, label=above:8]	(w4)	at (5,1)	{};
\path
\foreach \i/\j in {1/2,3/4}{
   	(w\i) edge [ev]   (b\j)
	(b\i) edge [ev]  node [c, label=above:$c$] {} (w\j)
    }
\foreach \i in {1,2,3,4}{
	(w\i) edge 	[ev]				(b\i)
	(w\i)	edge [ev,bend left=30]	(b\i)
	(w\i)	edge [ev,bend right=30]	(b\i)
	}
\foreach \i in {w1,b1,w2,b2}{
   	(\i) edge [e] (1)
	}
\foreach \i in {w3,b3,w4,b4}{
   	(\i) edge [e] (2)
	}	
\foreach \i/\j in {w2/b3,w3/b2}{
   	(\i) edge [es] (\j)
	};
}
\end{tikzpicture}
}

\newcommand{\cfish}{
\begin{tikzpicture}[x=2ex,y=2ex,baseline={([yshift=-.59ex]current bounding box.center)}] 
\scriptsize{
\node [v]	(1)	at (0,0)	{};
\node [v]	(2)	at (4,0)	{};
\node [vb]	(w1)	at (-1,-1)	{};
\node [vb]	(b1)	at (-1,1)	{};
\node [vb]	(w2)	at (1,1)	{};
\node [vb]	(b2)	at (1,-1)	{};
\node [vb]	(w3)	at (3,-1)	{};
\node [vb]	(b3)	at (3,1)	{};
\node [vb]	(w4)	at (5,1)	{};
\node [vb]	(b4)	at (5,-1)	{};
\path
\foreach \i/\j in {1/2,3/4}{
   	(w\i) edge [ev]   (b\j)
	(b\i) edge [ev]  node [c] {} (w\j)
	}
\foreach \i in {1,2,3,4}{
	(w\i) edge 	[ev]				(b\i)
	(w\i)	edge [ev,bend left=30]	(b\i)
	(w\i)	edge [ev,bend right=30]	(b\i)
	}
\foreach \i in {w1,b1,w2,b2}{
   	(\i) edge [e] (1)
	}
\foreach \i in {w3,b3,w4,b4}{
   	(\i) edge [e] (2)
	}	
\foreach \i/\j in {w2/b3,w3/b2}{
   	(\i) edge [es] (\j)
	};
}
\end{tikzpicture}
}

\newcommand{\cfisha}{
\begin{tikzpicture}[x=2ex,y=2ex,baseline={([yshift=-.59ex]current bounding box.center)}] 
\scriptsize{
\node [v]	(1)	at (0,0)	{};
\node [v]	(2)	at (4,0)	{};
\node [vb]	(w1)	at (-1,-1)	{};
\node [vb]	(b1)	at (-1,1)	{};
\node [vb]	(w2)	at (1,1)	{};
\node [vb]	(b2)	at (1,-1)	{};
\node [vb]	(w3)	at (3,-1)	{};
\node [vb]	(b3)	at (3,1)	{};
\node [vb]	(w4)	at (5,1)	{};
\node [vb]	(b4)	at (5,-1)	{};
\path
\foreach \i/\j in {1/2,3/4}{
   	(w\i) edge [ev]   (b\j)
	(b\i) edge [ev]  node [c] {} (w\j)
	}
\foreach \i in {1,2,3,4}{
	(w\i) edge 	[ev]				(b\i)
	(w\i)	edge [ev,bend left=30]	(b\i)
	(w\i)	edge [ev,bend right=30]	(b\i)
	}
\foreach \i in {w1,b1,w2,b2}{
   	(\i) edge [e] (1)
	}
\foreach \i in {w3,b3,w4,b4}{
   	(\i) edge [e] (2)
	}	
\foreach \i/\j in {w3/b2}{
   	(\i) edge [es] (\j)
	};
}
\end{tikzpicture}
}

\newcommand{\cfishb}{
\begin{tikzpicture}[x=2ex,y=2ex,baseline={([yshift=-.59ex]current bounding box.center)}] 
\scriptsize{
\node [v]	(1)	at (0,0)	{};
\node [v]	(2)	at (4,0)	{};
\node [vb]	(w1)	at (-1,-1)	{};
\node [vb]	(b1)	at (-1,1)	{};
\node [vb]	(w2)	at (1,1)	{};
\node [vb]	(b2)	at (1,-1)	{};
\node [vb]	(w3)	at (3,-1)	{};
\node [vb]	(b3)	at (3,1)	{};
\node [vb]	(w4)	at (5,1)	{};
\node [vb]	(b4)	at (5,-1)	{};
\path
\foreach \i/\j in {1/2,3/4}{
   	(w\i) edge [ev]   (b\j)
	(b\i) edge [ev]  node [c] {} (w\j)
	}
\foreach \i in {1,2,3,4}{
	(w\i) edge 	[ev]				(b\i)
	(w\i)	edge [ev,bend left=30]	(b\i)
	(w\i)	edge [ev,bend right=30]	(b\i)
	}
\foreach \i in {w1,b1,w2,b2}{
   	(\i) edge [e] (1)
	}
\foreach \i in {w3,b3,w4,b4}{
   	(\i) edge [e] (2)
	}	
\foreach \i/\j in {w2/b3}{
   	(\i) edge [es] (\j)
	};
}
\end{tikzpicture}
}

\newcommand{\cfishc}{
\begin{tikzpicture}[x=2ex,y=2ex,baseline={([yshift=-.59ex]current bounding box.center)}] 
\scriptsize{
\node [v]	(1)	at (0,0)	{};
\node [v]	(2)	at (4,0)	{};
\node [vb]	(w1)	at (-1,-1)	{};
\node [vb]	(b1)	at (-1,1)	{};
\node [vb]	(w2)	at (1,1)	{};
\node [vb]	(b2)	at (1,-1)	{};
\node [vb]	(w3)	at (3,-1)	{};
\node [vb]	(b3)	at (3,1)	{};
\node [vb]	(w4)	at (5,1)	{};
\node [vb]	(b4)	at (5,-1)	{};
\path
\foreach \i/\j in {1/2,3/4}{
   	(w\i) edge [ev]   (b\j)
	(b\i) edge [ev]  node [c] {} (w\j)
	}
\foreach \i in {1,2,3,4}{
	(w\i) edge 	[ev]				(b\i)
	(w\i)	edge [ev,bend left=30]	(b\i)
	(w\i)	edge [ev,bend right=30]	(b\i)
	}
\foreach \i in {w1,b1,w2,b2}{
   	(\i) edge [e] (1)
	}
\foreach \i in {w3,b3,w4,b4}{
   	(\i) edge [e] (2)
	};
}
\end{tikzpicture}
}

\newcommand{\cvsla}{
\begin{tikzpicture}[x=2ex,y=2ex,baseline={([yshift=0ex]current bounding box.center)}]
\scriptsize{
\node [v]	(1)	at (0,0)	{};
\begin{scope}[rotate=120]
\node [vb, label=left:1]		(b1)	at (.7,1.2)	{};
\node [vb, label=below:2]		(w1)	at (-.7,1.2)	{};
\end{scope}
\begin{scope}[rotate=240]
\node [vb, label=below:7]		(b2)	at (.7,1.2)	{};
\node [vb, label=right:8]		(w2)	at (-.7,1.2)	{};
\end{scope}
\begin{scope}[rotate=0]
\node [vb, label=right:5]		(b3)	at (.7,1.2)	{};
\node [vb, label=left:4]		(w3)	at (-.7,1.2)	{};
\end{scope}	
\path
(w1) edge [ev]  node [c, label=below:$c$] {} (b2)
\foreach \i/\j in {2/3,3/1}{
   	(w\i) edge [ev]   (b\j)
	}
\foreach \i in {1,2,3}{
	(w\i) edge 	[ev]				(b\i)
	(w\i)	edge [ev,bend left=30]	(b\i)
	(w\i)	edge [ev,bend right=30]	(b\i)
	}
\foreach \i in {w1,b1,w2,b2,w3,b3}{
   	(\i) edge [e] (1)
	}
(w3) edge [es, bend left=90] (b3);
}
\end{tikzpicture}
}

\newcommand{\cvslb}{
\begin{tikzpicture}[x=2ex,y=2ex,baseline={([yshift=-1.2ex]current bounding box.center)}]
\scriptsize{
\node [v]	(1)	at (0,0)	{};
\begin{scope}[rotate=60]
\node [vb, label=above:1]		(b1)	at (.7,1.2)	{};
\node [vb, label=left:2]		(w1)	at (-.7,1.2)	{};
\end{scope}
\begin{scope}[rotate=180]
\node [vb, label=left:3]		(b2)	at (.7,1.2)	{};
\node [vb, label=right:6]		(w2)	at (-.7,1.2)	{};
\end{scope}
\begin{scope}[rotate=300]
\node [vb, label=right:7]		(b3)	at (.7,1.2)	{};
\node [vb, label=above:8]		(w3)	at (-.7,1.2)	{};
\end{scope}	
\path
(b1) edge [ev]  node [c, label=above:$c$] {} (w3)
\foreach \i/\j in {1/2,2/3}{
   	(w\i) edge [ev]   (b\j)
	}
\foreach \i in {1,2,3}{
	(w\i) edge 	[ev]				node 	{}	(b\i)
	(w\i)	edge [ev,bend left=30]	node 	{}	(b\i)
	(w\i)	edge [ev,bend right=30]	node 	{}	(b\i)
	}
\foreach \i in {w1,b1,w2,b2,w3,b3}{
   	(\i) edge [e] (1)
	}
(w2) edge [es, bend left=90] (b2);
}
\end{tikzpicture}
}

\newcommand{\cvslc}{
\begin{tikzpicture}[x=2ex,y=2ex,baseline={([yshift=-.59ex]current bounding box.center)}]
\scriptsize{
\node [v]	(1)	at (0,0)	{};
\begin{scope}[rotate=90]
\node [vb, label=left:1]		(b1)	at (.7,1.2)	{};
\node [vb, label=left:2]		(w1)	at (-.7,1.2)	{};
\end{scope}
\begin{scope}[rotate=210]
\node [vb, label=below:5]		(b2)	at (.7,1.2)	{};
\node [vb, label=right:8]		(w3)	at (-.7,1.2)	{};
\end{scope}
\begin{scope}[rotate=330]
\node [vb, label=right:7]		(b3)	at (.7,1.2)	{};
\node [vb, label=above:4]		(w2)	at (-.7,1.2)	{};
\end{scope}	
\path
(b1) edge [ev]  node [c, label=150:$c_1$] {} (w2)
(w2) edge [ev]  node [c, label=30:$c_2$] {} (b3)
\foreach \i/\j in {1/1,1/2,3/2,3/3}{
   	(w\i) edge [ev]   (b\j)
	}
\foreach \i in {1,2,3}{
	(w\i)	edge [ev,bend left=30]	node 	{}	(b\i)
	(w\i)	edge [ev,bend right=30]	node 	{}	(b\i)
	}
\foreach \i in {w1,b1,w2,b2,w3,b3}{
   	(\i) edge [e] (1)
	}
(w2) edge [es, bend left=140] (b2);
}
\end{tikzpicture}
}

\newcommand{\cvsld}{
\begin{tikzpicture}[x=2ex,y=2ex,baseline={([yshift=-.59ex]current bounding box.center)}]
\scriptsize{
\node [v]	(1)	at (0,0)	{};
\begin{scope}[rotate=90]
\node [vb, label=left:1]		(b1)	at (.7,1.2)	{};
\node [vb, label=left:2]		(w1)	at (-.7,1.2)	{};
\end{scope}
\begin{scope}[rotate=210]
\node [vb, label=below:6]		(b2)	at (.7,1.2)	{};
\node [vb, label=right:8]		(w3)	at (-.7,1.2)	{};
\end{scope}
\begin{scope}[rotate=330]
\node [vb, label=right:7]		(b3)	at (.7,1.2)	{};
\node [vb, label=above:3]		(w2)	at (-.7,1.2)	{};
\end{scope}	
\path
(b1) edge [ev]  node [c, label=150:$c_1$] {} (w2)
(w2) edge [ev]  node [c, label=30:$c_2$] {} (b3)
\foreach \i/\j in {1/1,1/2,3/2,3/3}{
   	(w\i) edge [ev]   (b\j)
	}
\foreach \i in {1,2,3}{
	(w\i)	edge [ev,bend left=30]	node 	{}	(b\i)
	(w\i)	edge [ev,bend right=30]	node 	{}	(b\i)
	}
\foreach \i in {w1,b1,w2,b2,w3,b3}{
   	(\i) edge [e] (1)
	}
(w2) edge [es, bend left=140] (b2);
}
\end{tikzpicture}
}

\newcommand{\cvsd}{
\begin{tikzpicture}[x=2ex,y=2ex,baseline={([yshift=-.59ex]current bounding box.center)}]
\scriptsize{
\node [v]	(1)	at (0,0)	{};
\begin{scope}[rotate=90]
\node [vb]		(b1)	at (.7,1.2)	{};
\node [vb]		(w1)	at (-.7,1.2)	{};
\end{scope}
\begin{scope}[rotate=210]
\node [vb]		(b2)	at (.7,1.2)	{};
\node [vb]		(w3)	at (-.7,1.2)	{};
\end{scope}
\begin{scope}[rotate=330]
\node [vb]		(b3)	at (.7,1.2)	{};
\node [vb]		(w2)	at (-.7,1.2)	{};
\end{scope}	
\path
\foreach \i/\j in {1/1,1/2,2/1,2/3,3/2,3/3}{
   	(w\i) edge [ev]   (b\j)
	}
\foreach \i in {1,2,3}{
	(w\i)	edge [ev,bend left=30]	node 	{}	(b\i)
	(w\i)	edge [ev,bend right=30]	node 	{}	(b\i)
	}
\foreach \i in {w1,b1,w2,b2,w3,b3}{
   	(\i) edge [e] (1)
	}
(w2) edge [es, bend left=140] (b2);
}
\end{tikzpicture}
}